\documentclass[aps,prl,groupedaddress,twocolumn]{revtex4-1}

\usepackage{xcolor}
\usepackage{epsfig}
\usepackage{relsize}
\usepackage{graphicx}
\usepackage{amsfonts}
\usepackage[figuresright]{rotating}
\usepackage{amssymb}
\usepackage{amsmath}
\usepackage{color}
\usepackage{graphicx,bm,units,yfonts}

\usepackage{graphicx}
\DeclareGraphicsExtensions{.eps, .PNG, .jpg}

\newcommand{\ket}[1]{| #1 \rangle}
\newcommand{\bra}[1]{\langle #1 |}

\begin{document}

\title{Robust topological invariants of topological crystalline phases in the presence of impurities}
\author{Ian Mondragon-Shem$^{1}$ and Taylor L. Hughes$^{2}$}
\affiliation{$^1$ Department of Physics, Yale University, New Haven, Connecticut 06520, USA}
\affiliation{$^2$ Department of Physics, University of Illinois-Urbana Champaign, Urbana, IL 61801, USA}
\date{\today}
\begin{abstract}
Topological crystalline phases (TCPs) are topological states protected by spatial symmetries. A broad range of TCPs have been conventionally studied by formulating topological invariants (symmetry indicators) at invariant momenta in the Brillouin zone, which leaves open the question of their stability in the absence of translational invariance. In this work, we show that robust basis-independent topological invariants can be generically constructed for TCPs using projected symmetry operators.  Remarkably we show that the real-space topological markers of these invariants are exponentially localized to the fixed points of the spatial symmetry. As a result, this real-space structure  protects them against the presence of impurities that are located away from the fixed points. By considering all possible symmetry centers in a crystalline system we can generate a mesh of real-space topological markers that can provide a local topological distinction for TCPs. We illustrate the  robustness of this mesh of invariants with 1D and 2D TCPs protected by inversion, rotational and mirror symmetries. Finally, we find that the boundary modes of these TCPs can also exhibit robust topological invariants with localized markers on the edges. We illustrate this with the gapless Majorana boundary modes of mirror-symmetric topological superconductors, and relate their integer topological edge invariant with a quantized effective edge polarization.
\end{abstract}

\maketitle

%%%%%%%%%%%%%%%%%%%%%%%%%%%%%%%%%%%%%%%%%%%%%%%%%%%%%%%%%%%%%%%%%%%%%%%%%%%%%%%%%%%%%%%%%%%%%%%

\textit{Introduction.} Over the past decade, dramatic progress has been made in understanding the interplay between symmetry and topology in  quantum phases of matter \cite{Hasan2010, Qi2011}.
A particular focus has been the so-called topological crystalline phases (TCP) \cite{Fu2011,Hsieh2012}, which are topological states protected by spatial symmetries. A number of experiments have already revealed their existence in nature \cite{Xu2012,Dziawa2012,Tanaka2012,Okada2013}, and over the course of just a few years, significant theoretical progress has been made \cite{Hughes2011,Fang2012, Chiu2013, Zhang2013,  Fang2013a, Alexandradinata2014, Teo2013, Benalcazar2014,Yao2013, Isobe2015, Qi2015, Chiu2015, Lapa2016, Song2016, Bradlyn2017, Cano2018, Benalcazar61, Benalcazar2017, Schindlereaat0346}.

It has been argued that the surface states of TCPs are robust to the presence of disorder as long as the disorder is symmetric on average\cite{Fu2012,Fulga2014,Song2015}. 
Indeed experimental signatures of TCPs appear to be qualitatively consistent with the notion of an average symmetry, even in crystals that are expected to have moderate amounts of disorder from alloying, e.g., Pb$_x$Sn$_{1-x}$Te\cite{Zeljkovic2015}. Bulk studies have also argued the existence of statistical topological insulators which have protected boundary modes as long as the spatial symmetry is preserved on average \cite{Fulga2014,Diez2015,Song2015}.  Notwithstanding these studies, a bulk understanding of the mechanism behind the robustness of TCPs against the presence of impurities is still lacking, especially in cases in which a notion of average symmetry cannot be clearly defined.  

In this work, we find that a large class of TCPs can be characterized by topological invariants that are robustly quantized in the presence of impurities. To this end, we construct basis-independent topological invariants of TCPs with point-group symmetry. We argue that, since these invariants are expressed in terms of projected symmetry operators, their topological markers are \emph{spatially localized} to the fixed points of the spatial symmetry. As a result, such invariants are robust againts the presence of impurities. We exemplify this for 1D inversion-symmetric insulators and 2D rotation and mirror-symmetric topological superconductors. In the latter case, we also find phases in which the edge states are characterized by topological invariants that are themselves robustly quantized in the presence of impurities.

%%%%%%%%%%%%%%%%%%%%%%%%%%%%%%%%%%%%%%%%%%%%%%%%%%%%%%%%%%%%%%%%%%%%%%%%%%%%%%%%%%%%%%%%%%%%%%%

\textit{Case studies of localized topological markers.} We begin by analyzing 1D topological insulators protected by inversion symmetry with an even number of sites $L_x.$  On a system with periodic boundary conditions and lattice translation symmetry, there is an extensive set of possible inversion symmetry operations $P_{\mathcal{S}}$ that each leave two positions on the lattice invariant: $\mathcal{S}=\{R_1+r_0, R_2+r_0\},$  where $R_{1,2}$ label unit cells, and $r_0$ takes one of two values $0,1/2.$ We thus consider a gapped Hamiltonian $\hat{H}$ such that $[\hat{p}_{\mathcal{S}},\hat{H}]=0,$ where $\hat{p}_{\mathcal{S}}$ is the Hilbert space representation of $P_{\mathcal{S}}.$ 

The conventional topological classification of inversion symmetric insulators proceeds by formulating the problem in momentum space $\hat{H}=\sum_{k_x} c^{\dagger}_{k_x} h(k_x) c^{\phantom{\dagger}}_{k_x}.$ The Bloch Hamiltonian must satisfy $\widetilde{p}_{\mathcal{S}}(k_x) h(k_x) \widetilde{p}^{-1}_{\mathcal{S}}(k_x)=h(P_{\mathcal{S}}k_x),$ where  $\widetilde{p}_{\mathcal{S}}(k_x)$ is the momentum representation of the inversion operator. At the inversion-invariant momenta $k^{\text{inv}}_{k_x}=0,\pi,$ Bloch states are labeled by inversion eigenvalues. Let  $n_{\mathcal{S}}^{(\pm)}(k_{x}^{\text{inv}})$ denote the number of occupied states with inversion eigenvalues $\pm 1$ at momentum $k_{x}^\text{inv}.$ Let us define the topological invariant $\Delta^p_{\mathcal{S}}=\sum_{k_{x}^{\text{inv}}}\left[n_{\mathcal{S}}^{(+)}(k_{x}^{\text{inv}})-n_{\mathcal{S}}^{(-)}(k_{x}^{\text{inv}})\right].$ In  \cite{Fang2013b},  it was shown that $\Delta^p_{\mathcal{S}}\vert_{r_0=1/2}$ is the integer invariant that distinguishes inversion-symmetric topological states from the trivial atomic limit. A simple model with nontrivial $\Delta^p_{\mathcal{S}}\vert_{r_0=1/2}$ is $h(k_x)=\sin{k_x} \sigma_2+(m-\cos{k_x}) \sigma_1,$ where the $\sigma_{a}$ are Pauli matrices acting on two orbitals $\{A,B\}$ per cell. The inversion operator is $\widetilde{p}_{\mathcal{S}}(k_{x}^{\text{inv}})=e^{-2i k_{x}^{\text{inv}} r_0} \sigma_1,$ which leads to $\Delta^p_{\mathcal{S}}\vert_{r_0=1/2}=2 (0)$ in the topological (trivial) phase.

\begin{figure}
\centering
\includegraphics[scale=0.19]{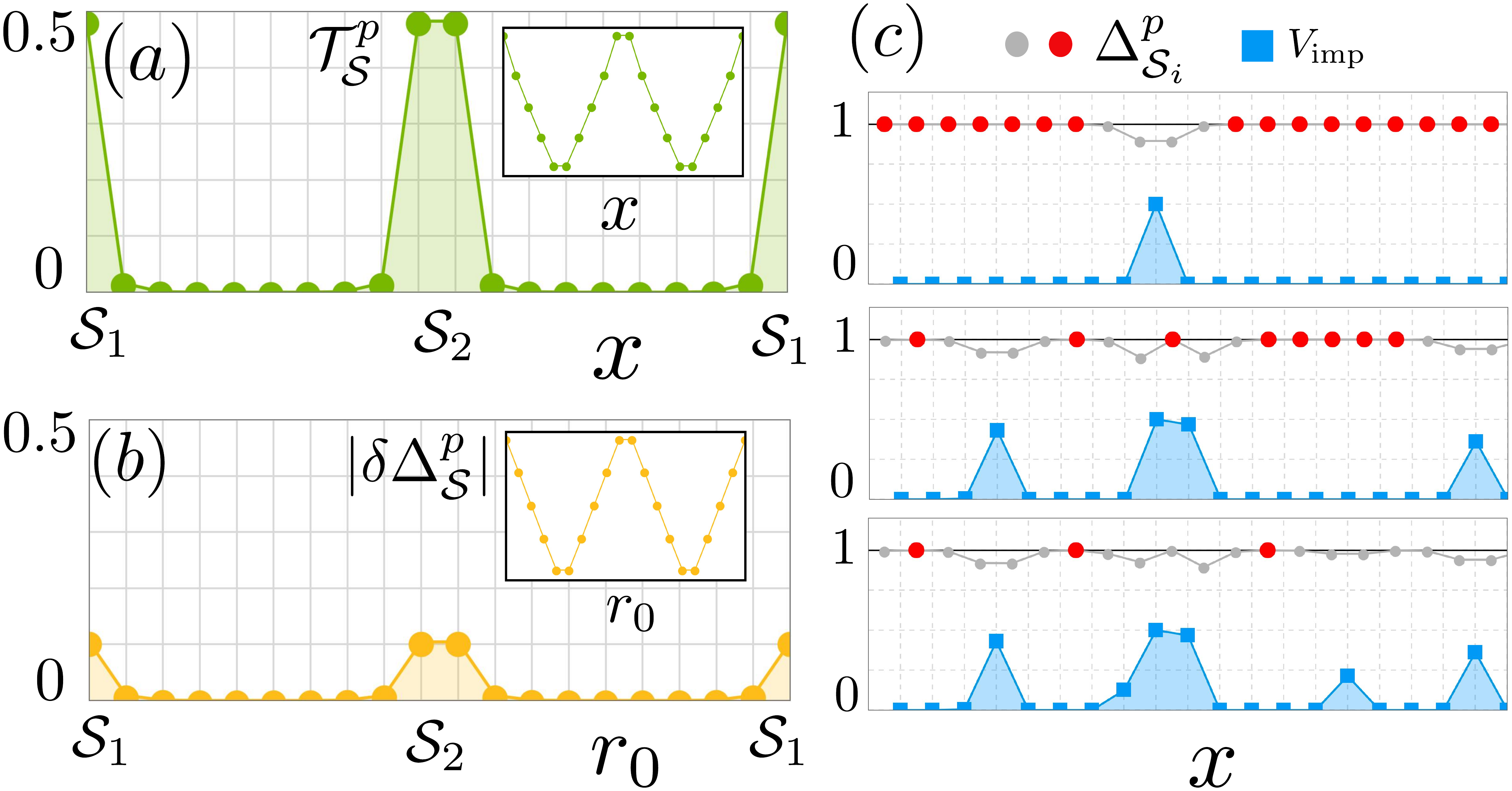}
\caption{Inversion-symmetric 1D insulator. \textbf{(a)} Topological marker $\mathcal{T}^{p}_{\mathcal{S}}(x).$ The inset shows the same data on a logarithmic scale. \textbf{(b)} Deviation of the invariant $\vert\delta \Delta^{p}_{\mathcal{S}}\vert$ induced by an impurity at position $X_0.$ The inset shows the same data on a logarithmic scale. \textbf{(c)} Sequence of plots showing $ \Delta^{p}_{\mathcal{S}_i} $ (dots) for all $\mathcal{S}$ in a system with random impurities (blue squares). From top to bottom, the density of impurities increases. The dots that are red denote the cases for which $\vert \Delta^{p}_{\mathcal{S}_i}\vert$ differs from the clean-limit value $1.0$ by $10^{-3},$ whereas gray dots deviate more significantly.}\label{Fig_inv}
\end{figure}

As it stands, the computation of $\Delta^p_{\mathcal{S}}$ relies on momentum being a good quantum number. It is thus unclear to what extent $\Delta^p_{\mathcal{S}}$ can remain quantized when translational symmetry is lost due to the presence of impurites in the system. Even with translation symmetry, using momentum states is just a choice of basis. There must necessarily be a way to compute the topological invariant in a basis-independent manner. To see that this is indeed possible, we define the projected inversion operator $\bar{p}_{\mathcal{S}}= \mathcal{P}_{\text{G}}\hat{p}_{\mathcal{S}}\mathcal{P}_G,$ where $\mathcal{P}_G=\sum_{n\in \text{occ.}} \ket{u_n}\bra{u_n}$ projects into the occupied states. We find that 
\begin{equation}
\Delta^p_{\mathcal{S}}=\text{Tr}[\bar{p}_{\mathcal{S}}].\label{Eq_InvInv}
\end{equation}
This basis-independent expression liberates us from the momentum-space description. In particular, instead of analyzing the contributions to $\Delta^p_{\mathcal{S}}$ in momentum space, we can now determine its real-space behavior. It is revealing in this regard to examine the spatially-resolved topological marker $\mathcal{T}^p_{\mathcal{S}}(x)= \bra{x}\text{Tr}_\circ[\bar{p}_{\mathcal{S}}]\ket{x},$ where $\text{Tr}_\circ$ traces over the local degrees of freedom in each unit cell, and $\Delta^p_{\mathcal{S}}=\sum_{x=1}^{L_x}\mathcal{T}^p_{\mathcal{S}}(x).$ Topological markers have been used recently in other contexts to study the local properties of topological phases \cite{Bianco2011,Meier929}. Since  $\bra{x,\alpha} \hat{p}_{\mathcal{S}} \ket{x',\beta} \propto \delta_{x,P_{\mathcal{S}}x'},$ and $\vert \bra{x',\alpha} \mathcal{P}_G \ket{x, \beta} \vert < \mathcal{O}\left( e^{-\vert x-x' \vert/\zeta}\right)$ when  $\vert x-x' \vert \gg\zeta,$ we find that $\mathcal{T}^p_{\mathcal{S}}(x)$ satisfies
\begin{equation}
\vert \mathcal{T}^p_{\mathcal{S}}(x)\vert < \mathcal{O}\left( e^{-2\vert x-\mathcal{S} \vert/\zeta}\right), \quad  \vert x-\mathcal{S} \vert\gg \zeta, \label{Eq_TInv}
\end{equation}
where $\vert x-\mathcal{S}\vert$ denotes the smallest distance of $x$ to any of the points in $\mathcal{S},$ and the length scale $\zeta$ is roughly of the order of the inverse gap/localization length. By taking into account the full set of operators $P_{\mathcal{S}}$ for a translationally invariant lattice, the topology of the inversion-symmetric insulator is then comprised of a real-space mesh of localized topological markers. We illustrate one of these localized markers in Fig.\ref{Fig_inv}a; the inset shows it on a logarithmic scale to confirm its exponential localization.

The localized nature of the marker $\mathcal{T}^p_{\mathcal{S}}(x)$ protects the invariant $\Delta^p_{\mathcal{S}}$ against the presence of impurities. To see this, suppose we add an impurity at a position $X_0.$ The projection operator of the new ground state can be written as $\mathcal{P}'_G=\mathcal{P}_G+\delta \mathcal{P}_G.$ Since the system is gapped, $\delta \mathcal{P}_G$ only has support in an exponentially small neighborhood of size $\zeta$ centered at the impurity. For an impurity placed far away from $\mathcal{S},$ and using $\bra{x}\delta \mathcal{P}_G\ket{X_0}< \mathcal{O}\left(e^{-\vert X_0-x\vert/\zeta}\right),$ we find that
\begin{equation}   \vert\delta\Delta^{p}_{\mathcal{S}}\vert\equiv\vert \Delta^{p'}_{\mathcal{S}}(X_0)-\Delta^{p}_{\mathcal{S}}\vert<\mathcal{O}\left(e^{-2\vert X_0 -\mathcal{S}\vert/\zeta}\right).
\end{equation}
Thus, the effect of the impurity on the invariant is exponentially suppressed in the distance between the impurity center and the inversion center(s). To confirm the robustness of $\Delta^{p}_{\mathcal{S}},$ in Fig.\ref{Fig_inv}b we show $\vert\delta\Delta^{p}_{\mathcal{S}}\vert$ as a function of $X_0$ (the inset shows its behavior on a logarithmic scale). It is clear that the change in the invariant is exponentially suppressed as the impurity moves away from $\mathcal{S}$ because correlations in the insulator are short-ranged. As a result, $\Delta^{p}_{\mathcal{S}}$ remains quantized when $X_0$ is further than $\zeta$ from $\mathcal{S}.$ We show the effect of progressively adding more impurities in Fig.\ref{Fig_inv}c. Here we calculate the topological marker integrated over a neighborhood of \textit{one} inversion center: $\Delta^{p}_{\mathcal{S}_i}=\sum_{\vert x-\mathcal{S}_i\vert  < \zeta}\mathcal{T}^{p}_{\mathcal{S}}(x),$  such that  $\Delta^p_\mathcal{S}\approx\sum_{i=1,2}\Delta^{p}_{\mathcal{S}_i},$ where $\mathcal{S}_i=R_i+r_0$ runs over the two invariant points in $\mathcal{S}.$ For our example model in the topological phase, and without impurities, $\Delta^{p}_{\mathcal{S}_i}\vert_{r_0=1/2}=1,$ hence $\Delta^{p}_{\mathcal{S}}\vert_{r_0=1/2}=2.$ Fig.\ref{Fig_inv}c shows $\Delta^{p}_{\mathcal{S}_i}$ for the mesh of all possible $\mathcal{S}$ as impurities (blue squares) are progressively added to the system. Without impurities we find a uniformly quantized mesh of topological invariants. The impurities destroy the quantization of the mesh at some points (gray dots), while regions subsist in which $\Delta^{p}_{\mathcal{S}_i}$ is robustly quantized (red dots). Thus, we find that even in the presence of impurities the system can retain a local distinction between trivial and non-trivial topological phases (atomic limits).
%%%%%%%%%%%%%%%%%%%%%%%%%%%%%%%%%%%%%%%%%%%%%%%%%%%%%%%%%%%%%%%%%%%%%%%%%%%%%%%%%%%%%%%%%%%%%%%

\begin{figure}
\begin{center}
\includegraphics[trim=0cm 0cm 0cm 0cm,scale=0.23]{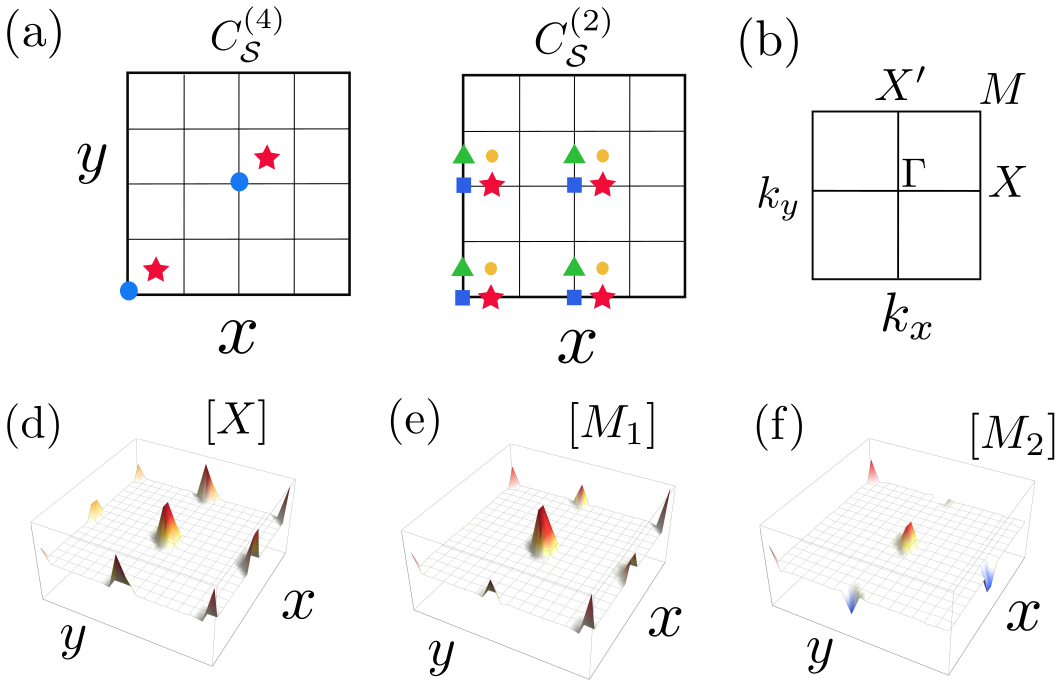}
\caption{Rotationally-symmetric 2D superconductors. \textbf{(a)} Invariant points under $C^{(4)}_{\mathcal{S}}$ (left) and $C^{(2)}_{\mathcal{S}}$ (right). For example, in the former case, each set of symbols maps to themselves under four-fold rotations centered at that symbol. \textbf{(b)} Invariant momenta in the BZ. \textbf{(c,d,e)} Topological markers for the rotationally symmetric topological superconductor with $(\text{Ch}, [X],[M_1],[M_2])=(1,1,1,0).$  }\label{Fig_Rot}
\end{center}
\end{figure}

Next, we consider 2D topological superconductors protected by rotation symmetries and without time-reversal symmetry. With periodic boundary conditions, a given $n$-fold rotation operation $C^{(n)}_{\mathcal{S}}$ leaves invariant a set $\mathcal{S}=\{\textbf{R}_i+\mathbf{r}_0\}$ of positions on the lattice (see Fig. \ref{Fig_Rot}a).  In particular, with primitive vectors $2\textbf{a}_{1}$  and $2\textbf{a}_{2},$ then $C^{(4)}_{\mathcal{S}}$ allows $\mathbf{r}_0\in \{\textbf{0}, \textbf{a}_1+\textbf{a}_2\},$ whereas $C^{(2)}_{\mathcal{S}}$ allows $\textbf{r}_0\in\{\textbf{0},\textbf{a}_1,\textbf{a}_2, \textbf{a}_1+\textbf{a}_2\}.$ We consider models that satisfy $[\hat{c}^{(n=2,4)}_{\mathcal{S}},\hat{H}]=0$, where $\hat{c}^{(n)}_{\mathcal{S}}$ is the Hilbert space representation of $C^{(n)}_{\mathcal{S}}.$ The conventional classification consists of studying Bloch states at the rotation-invariant momenta shown in Fig.\ref{Fig_Rot}b: $\hat{c}^{(n=4)}_{\mathcal{S}}$ ($\hat{c}^{(n=2)}_{\mathcal{S}}$) leaves invariant the momenta $\Gamma, M$ ($\Gamma, M, X, X'$). At these momenta, the Bloch states can be labeled by rotation eigenvalues. One of the topological invariants of this TCP is a Chern number Ch, which exists regardless of the spatial symmetries. Additionally, there are three more integers $([X],[M_1],[M_2])$ \cite{Teo2013, Benalcazar2014}. These invariants count the number of occupied states at the $X,M$ points that have rotation eigenvalues $(e^{i \pi/2}, e^{i\pi/4}, e^{i 3 \pi/4})$ respectively,  relative to those at $\Gamma.$ 

Rather than writing out momentum-space expressions for these  invariants, we can again express them in a basis independent manner. By defining the projected rotation operators $\overline{c}^{(n)}_{\mathcal{S}}=\mathcal{P}_G\hat{c}^{(n)}_{\mathcal{S}}\mathcal{P}_G,$ we obtain
\begin{eqnarray}
[X]&=&\frac{i}{4}\text{Tr}\left[\left(\overline{c}^{(2)}_{\mathcal{S}}|_{\textbf{r}_0=\mathbf{a}_1}\right)+\left(\bar{c}^{(2)}_{\mathcal{S}}|_{\textbf{r}_0=\textbf{a}_1+\textbf{a}_2}\right)\right],\\
\left[M_{1,2}\right]&=&\frac{i}{4}\text{Tr}\left[\sqrt{2}\left(\bar{c}^{(4)}_{\mathcal{S}}|_{\textbf{r}_0=\mathbf{a}_1+\textbf{a}_2}\right)\pm \left(\bar{c}^{(2)}_{\mathcal{S}}|_{\textbf{r}_0=\mathbf{a}_1}\right)\right].
\end{eqnarray}
The structure of $\overline{c}^{(n)}_{\mathcal{S}}$ again leads to exponentially localized topological markers.  We exemplify this in Fig.\ref{Fig_Rot}d,e,f for a model having  $(\text{Ch}, [X],[M_1],[M_2])=(1,1,1,0)$  (model details in SM). Furthermore, similar to the inversion-symmetric case, this localization grants these TCPs with robustness against the addition of impurities.

Both the inversion and rotation symmetric examples suggest that other TCPs could also encode topological invariants that are robust to impurities. Naively, one might have attempted a systematic study of such robust invariants in terms of exponentially localized Wannier functions.  However, the rotational case shows that such topological markers are localized even in the presence of a nonzero Chern number, which precludes the construction of localized Wannier functions. The localized nature of topological markers is more general than a Wannier function description might allow. It is fundamentally due to the action of projected symmetry operators.

\textit{Robust topological invariants of TCPs.} We now show that robust invariants can indeed be constructed using projected symmetry operators for a large class of TCPs invariant under a spatial symmetry $\hat{g}_{\mathcal{S}}.$  Consider a Bloch Hamiltonian $h(\textbf{k}_G)$ of a $d$-dimensional system obtained by Fourier transforming a real-space Hamiltonian where \textit{only} the spatial components $\textbf{r}_G$ that change under $G_{\mathcal{S}}$ are transformed; we denote by $d_G (\leq d)$ the number of components of $\textbf{r}_G.$ At the invariant momenta, the states $\{\ket{u^n_{j\textbf{k}_G^{\text{inv}}}}\}$ are labeled by the eigenvalues $\{e^{i\phi_j(\textbf{k}_G^\text{inv})}\}$ of $\widetilde{g}_{\mathcal{S}}(\textbf{k}_G^{\text{inv}})$ where $j$ runs over the set of symmetry representations. 

The topology of a large class of TCPs is encoded by the projection operator at the invariant momenta  $\mathcal{P}_{j\textbf{k}_G^{\text{inv}}}=\sum_{n\in \text{occ.}}\ket{u^n_{j\textbf{k}_G^{\text{inv}}}}\bra{u^n_{j\textbf{k}_G^{\text{inv}}}}.$ Let us generically denote the topological invariant encoded by  $\mathcal{P}_{j\textbf{k}_G^{\text{inv}}}$ as $\tau_j(\mathbf{k}_G^{\text{inv}}).$ As examples, $\tau_j(\mathbf{k}_G^{\text{inv}})$ could be the number of occupied states with a particular rotation eigenvalue, or it could be a mirror-Chern number \cite{Fu2011}, or a mirror-winding  number \cite{Zhang2013}. In all such examples, the $\tau_j(\mathbf{k}_G^{\text{inv}})$ can be written in the generic form $\tau_j(\mathbf{k}_G^{\text{inv}})=\text{Tr}[\mathcal{F}(\mathcal{P}_{j\mathbf{k}_G^{\text{inv}}})],$ where $\mathcal{F}$ is a function of $\mathcal{P}_{j\mathbf{k}_G^{\text{inv}}}$, with the constraint that it ultimately projects into a sum over occupied states.  For example, to count the occupied rotation and inversion representations we just have  $\mathcal{F}(\mathcal{P}_{j\textbf{k}^{\text{inv}}_{G}})\propto\mathcal{P}_{j\mathbf{k}_G^{\text{inv}}}$ where  ($d_G=d$ in 2D). More generally, it could also involve commutators of $\mathcal{P}_{j\textbf{k}_G^{\text{inv}}}$ with position operators that are \textit{not} components of $\textbf{r}_G.$ Examples include mirror ($y\rightarrow -y$) winding numbers for which  $\mathcal{F}(\mathcal{P}_{jk_y^{\text{inv}}})\propto \mathcal{P}_{jk_y^{\text{inv}}}[\hat{X},\mathcal{P}_{jk_y^{\text{inv}}}]$ where $(d=2,d_G=1);$ and mirror ($z\to-z$) Chern invariants for which $\mathcal{F}(\mathcal{P}_{jk_z^{\text{inv}}})\propto \mathcal{P}_{j k_z^{\text{inv}}}[[\hat{X},\mathcal{P}_{jk_z^{\text{inv}}}], [\hat{Y},\mathcal{P}_{jk_z^{\text{inv}}}]]$ where  $(d=3,d_G=1).$

Using these types of topological invariants, we can construct new robust invariants through a remarkably simple linear combination:
\begin{equation}
    T^{g}_{\mathcal{S}}\equiv\sum_{j,\textbf{k}_G^{\text{inv}}}  e^{i\phi_j(\textbf{k}_G^{\text{inv}})}  \tau_j(\textbf{k}_G^{\text{inv}}).\label{Eq_TgS}
\end{equation} 
By rewriting this expression in terms of a trace over the full Hilbert space, we find that $T^{g}_{\mathcal{S}}=\text{Tr}[\overline{g}_{\mathcal{S}} \mathcal{F}(\mathcal{P}_{\text{G}})],$  where $\overline{g}_{\mathcal{S}}=\mathcal{P}_G \hat{g}_{\mathcal{S}}\mathcal{P}_G,$ with  $\mathcal{P}_G=\sum_{j\mathbf{k}_G}\mathcal{P}_{j\mathbf{k}_G}$ (see SM for details).  The presence of $\overline{g}_{\mathcal{S}}$ leads to spatial topological markers that are localized in the vicinity of $\mathcal{S}$, i.e.,  $\vert\mathcal{T}^{g}_{\mathcal{S}}(\textbf{r}_G)\vert <\mathcal{O}\left(e^{-\vert G_{\mathcal{S}}\textbf{r}_G-\textbf{r}_{G}\vert/\zeta}\right) \quad \text{when} \quad \zeta \ll \vert G_{\mathcal{S}}\textbf{r}_G-\textbf{r}_{G}\vert.$ Furthermore, if we place an impurity at $\textbf{R}_0$ far from $\mathcal{S,}$ then $\vert T^{g'}_{\mathcal{S}}(\textbf{R}_0)-T^{g}_{\mathcal{S}}\vert<\mathcal{O}\left( e^{-\vert G_{\mathcal{S}}\textbf{R}_0- \textbf{R}_0\vert/\zeta}\right).$ We thus see that the topology of these TCPs is generically encoded in localized meshes of topological markers that are robust against the presence of impurities. These localized markers are consistent with other real-space layered approaches \cite{Song2016,Huang2017,Zhida2018}, as well as with approaches using path-integral formulations that use partial symmetry operators \cite{Shiozaki2017,Shapourian2017,Shiozaki2018}, though the focus of these works was not on the effects of disorder.

%%%%%%%%%%%%%%%%%%%%%%%%

\begin{figure}
\centering
\includegraphics[scale=0.19]{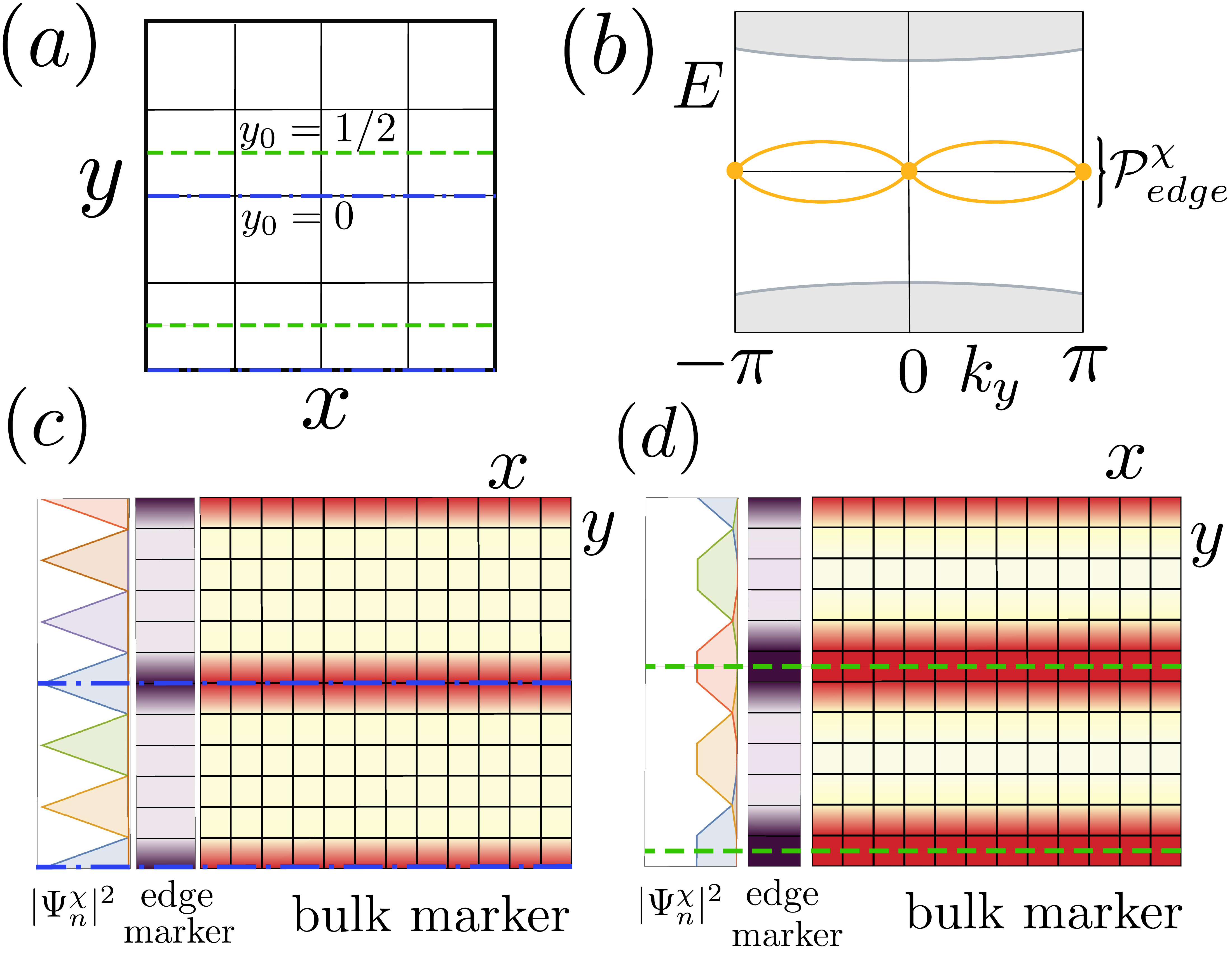}
\caption{Mirror-symmetric DIII 2D superconductor. \textbf{(a)} 
Mirror-invariant lines at $y_0=0$ (dash-dotted blue) and $y_0=1/2$ (dashed green).
\textbf{(b)} Schematic of the energy bands in the BZ for TCPs with trivial DIII strong index.  \textbf{(c,d)} Topological markers of the bulk (square lattice) and edge (purple strip) invariants, as well as some of the Wannier functions $\vert \Psi^{\chi}_n\vert^2$ at the edge for the phases (ii) and (iii) discussed in the main text. }\label{Fig_BulkWinding}
\end{figure}

\textit{Robust topological markers at the boundary.} The boundary modes of TCPs can also exhibit robust topological markers connected with the localized bulk invariants $T^{g}_{\mathcal{S}}.$ To exemplify this, consider 2D mirror-symmetric superconductors in class DIII \cite{Zhang2013}.  This symmetry class is described by a Bogoliubov-de Gennes Hamiltonian $h(k_y)$ that commutes with a mirror operator $\hat{m}_{\mathcal{S}}$ that sends $y\to -y,$  and where $\mathcal{S}=(Y_1+y_0,Y_2+y_0)$ denotes the mirror-invariant lines (see Fig.\ref{Fig_BulkWinding}a); we assume $\hat{m}^2_{\mathcal{S}}=-1.$ A chiral symmetry $\hat{\chi}=TC$ arises from having both time-reversal $T$ and particle-hole $C$ symmetries, and satisfies $\{T,\hat{\chi}\}=0.$ This implies that, in the basis in which $\hat{m}_{\mathcal{S}}$ is diagonal, each mirror block of $h(k^{\text{inv}}_y)$ belongs to class AIII \cite{Zhang2013}. Using the basis-independent topological winding invariants of class AIII  \cite{MS2014}, each mirror block has a topological winding invariant  $\nu^{\pm}_{\mathcal{S}}(k^{\text{inv}}_y)= \text{Tr}\left[\mathcal{F}(\mathcal{P}_{\pm,k^{\text{inv}}_y})\right],$ where  $\mathcal{F}(\mathcal{P}_{\pm,k^{\text{inv}}_y})= 2 L_x^{-1} \mathcal{P}_{\pm,k^{\text{inv}}_y}[\hat{X},\mathcal{P}_{\pm,k^{\text{inv}}_y}]\hat{\chi},$ where $\hat{X}$ is the position operator. By implementing Eq.\ref{Eq_TgS}, we obtain the robust (bulk) invariant $T^{m}_{\mathcal{S}}=\text{Tr}\left[\overline{m}_{\mathcal{S}} \mathcal{F}(\mathcal{P}_G) \right].$ We define $\mathcal{V}^m_{\mathcal{S}}= -\frac{i}{2}T^{m}_{\mathcal{S}},$ with a factor to simplify later expressions.

The presence of $\overline{m}_{\mathcal{S}}$ in $\mathcal{V}^m_{\mathcal{S}}$ localizes its real space topological markers to mirror lines. Consider three representative cases. (i) If $\mathcal{V}^m_{\mathcal{S}}\vert_{y_0=0}=\mathcal{V}^m_{\mathcal{S}}\vert_{y_0=1/2}=1,$ then the markers have support on every possible mirror line. This implies a non-trivial (strong) DIII $\mathbb{Z}_2$ index, and a single Kramers' pair of propagating helical Majorana modes at the boundary. (ii) If $\mathcal{V}^m_{\mathcal{S}}\vert_{y_0=0}\neq 0$ and $\mathcal{V}^m_{\mathcal{S}}\vert_{y_0=1/2}=0,$ then non-vanishing markers are centered at $y_0=0.$ Such a phase has a trivial DIII $\mathbb{Z}_2$ index, and is adiabatically connected to the limit of decoupled DIII mirror-symmetric wires parallel to the $x$-axis and stacked in the $y$-direction; as such it is a weak topological superconductor. (iii) If $\mathcal{V}^m_{\mathcal{S}}\vert_{y_0=0}=0$ and $\mathcal{V}^m_{\mathcal{S}}\vert_{y_0=1/2}\neq 0,$ then its DIII strong invariant is trivial, but it has non-trivial markers that are localized on $y$-lines centered halfway between cells. Thus, while this is still adiabatically connected to a decoupled wire limit, the wires are shifted by half a unit cell in the $y$-direction compared to case (ii). Indeed this shift is not arbitrary and is the result of a quantized edge topological invariant discussed below.  We illustrate the markers for the latter two cases in Fig. \ref{Fig_BulkWinding}c,d using models provided in the SM. We see that the marker for case (ii) (Fig. \ref{Fig_BulkWinding}c) is localized around the center of the unit cell, while the marker for case (iii) (Fig. \ref{Fig_BulkWinding}d) is localized halfway between unit cells. 

Heuristically, in cases (ii, iii) each mirror line corresponds to a non-trivial 1D topological superconductor that must have Kramers' pairs of Majorana end modes on boundaries with normal vector $\pm\hat{x}$. In case (ii) such Majorana modes should arise along the edge at integer sites in the $y$-direction, while in (iii) we expect them to be displaced to the mid-point between integers. This suggests that both phases could be distinguished by the location of Majorana modes along the edge.

To quantify this intuition, we examine the distribution of Wannier centers inside a unit cell $\mathcal{R}$ at the edge where a mirror line terminates. First, note that with a trivial strong index, the edge spectrum can always be detached from the bulk bands while preserving the DIII and mirror symmetries (see Fig. \ref{Fig_BulkWinding}b); we denote by $\mathcal{P}_{edge}$ the projector into this spectrally detached Majorana band (DMB). Since $\mathcal{P}_{edge}$ includes positive and negative energies related by chiral symmetry, we can write it in terms of projectors into chiral subspaces $\mathcal{P}_{edge}=\sum_{\chi=\pm}\mathcal{P}^{\chi}_{edge}.$ The Wannier centers $\{e^{-2\pi i (R_y+\nu_{R_yj})/L_y}\}$ and eigenstates $\{\ket{\Psi^{\chi}_{R_yj}}\}$ of the DMB are then obtained by diagonalizing the periodic projected position operator $\mathcal{W}^{\chi}_{edge}=\mathcal{P}^{\chi}_{edge} e^{-2\pi i \hat{Y}/L_y} \mathcal{P}^{\chi}_{edge}$ \cite{Kivelson1982}; here, $R_y$ labels unit cells along the edge, and $j$ runs over the $N_b$ states per unit cell included in the DMB. Using these states, we can locally characterize the edge modes at $\mathcal{R}$ by the edge shift $\delta\mathcal{Y}^{\chi}_{edge}(\mathcal{R})=\left(\sum_{j}\nu_{\mathcal{R}j} \right)\text{mod}1,$ where $\delta\mathcal{Y}^{+}_{edge}(\mathcal{R})=\delta\mathcal{Y}^{-}_{edge}(\mathcal{R})$ due to time-reversal symmetry. Furthermore, due to mirror symmetry $\delta\mathcal{Y}^{\chi}_{edge}(\mathcal{R})=-\delta\mathcal{Y}^{\chi}_{edge}(\mathcal{R}) \mod 1.$ Thus, the edge shift must be quantized to either $\delta\mathcal{Y}^{\chi}_{edge}(\mathcal{R})=0$ or $1/2.$ 

Now, the value of $\delta\mathcal{Y}^{\chi}_{edge}(\mathcal{R})$ can be determined by the (parity of the) number of Wannier centers located at the symmetry centers in the unit cell at $\mathcal{R},$ and the mirror related cell $\mathcal{R}'.$ Hence, consider the pair of invariant points $S\vert_{y_0=1/2}=\{\mathcal{R}+1/2,\mathcal{R}'+1/2\}.$ The Wannier states centered at these two symmetry centers are eigenstates of $\hat{m}_{\mathcal{S}}\vert_{y_0=1/2},$ whereas the remaining Wannier states come in mirror-related pairs. It follows that we can write the equality (see SM): $\delta\mathcal{Y}^{\chi}_{edge}(\mathcal{R})=\left[-\frac{i}{4}\sum_{R_y j}\bra{\Psi^{\chi}_{R_y j}}(\hat{m}_{\mathcal{S}}\vert_{y_0=1/2})\ket{\Psi^{\chi}_{R_y j}}\right] \text{mod}\, 1.$ The sum can be replaced by a trace over the Hilbert space of the entire system to obtain  $\delta\mathcal{Y}^{\chi}_{edge}(\mathcal{R})=\left[\frac{1}{4}\Delta^m_{\mathcal{S}}\vert_{y_0=1/2}\right]\text{mod}\, 1,$ where $\Delta^{m }_{\mathcal{S}}\equiv -\frac{i}{2}\text{Tr}[(\mathcal{P}_{edge}\hat{ {m}}_\mathcal{S}\mathcal{P}_{edge})\chi].$  If we evaluate the edge invariant $\Delta^m_{\mathcal{S}}$ and use the connection between the bulk AIII winding invariant and the number of zero-energy boundary modes, then we find (see SM for detail) $\Delta^{m}_{\mathcal{S}}=\mathcal{V}^{m}_{\mathcal{S}}.$ The edge shift is then determined by the bulk invariant
\begin{equation}
    \delta\mathcal{Y}^{\chi}_{edge}(\mathcal{R})=\left(\frac{1}{4}\mathcal{V}^m_{\mathcal{S}}\vert_{y_0=1/2}\right)\text{mod}\, 1.
\end{equation}
We thus conclude that the two possible quantized edge shifts $(0, 1/2)$ correspond to the two phases (ii, iii) we discussed earlier. 

A remarkable by-product of this analysis reveals that the DMB is characterized by the integer topological invariant $\Delta^{m }_{\mathcal{S}}$ which must be equal to the bulk invariant $\mathcal{V}^{m}_{\mathcal{S}}$. Its topological marker is also exponentially localized at $S$ along the edge and is robust to the presence of impurities at the edge. In Fig. \ref{Fig_BulkWinding}c,d we illustrate the markers of $\mathcal{V}^m_{\mathcal{S}}$ (labeled bulk marker) and $\Delta^{m }_{\mathcal{S}}$ (labeled edge marker) as well as representative Wannier functions $\vert \Psi^{\chi}_n\vert^2$ at the edge for two cases: $(\mathcal{V}^m_{\mathcal{S}}\vert_{y_0=0},\mathcal{V}^m_{\mathcal{S}}\vert_{y_0=1/2})=(2,0)$ and $(\mathcal{V}^m_{\mathcal{S}}\vert_{y_0=0},\mathcal{V}^m_{\mathcal{S}}\vert_{y_0=1/2})=(0,2).$ There is a clear correspondence between the bulk and edge topological markers that are centered at the mirror-invariant lines. The Wannier functions at the edge are consistently centered at these mirror-invariant points as well.

%%%%%%%%%%%%%%%%%%%%%%%%%%%%%%%%%%%%%%%%%%%%%%%%%%%%%%%%%%%%%%%%%%%%%%%%%%%%%%%%%%%%%%%%%%%%%%%

\textit{Conclusions.} In this work we have found that a large class of TCPs are characterized by robustly quantized bulk and edge topological invariants that are constructed using projected symmetry operators. Our results constitute a starting point to further explore the real-space structure of other kinds of TCPs, for example those on non-symmorphic lattices. Furthermore, it suggests a pathway to study disordered systems with interactions by replacing single-particle projectors with Green's functions \cite{Wang2012}. Finally, our findings point to the possibility of experimentally studying TCPs through local probes; for example, in ultra-cold atomic systems, where the single-particle projection operator can be extracted from time-of-flight measurements \cite{Zhang2009}.

\textit{Acknowledgements}
We thank M. Hermele and S. Velury for useful conversations. IMS gratefully acknowledges support by the Yale Prize Postdoctoral Fellowship.  TLH thanks the US National Science Foundation (NSF) grants EFMA-1627184 (EFRI), DMR-1351895 (CAREER) for support.

\clearpage

\onecolumngrid

\appendix

%%%%%%%%%%%%%%%%%%%%%%%%%%%%%%%%%%%%%%%%%%%%%%%

\begin{center}
\begin{Large}
\textbf{Supplementary Material}
\end{Large}   
\end{center}

\section{Section I: Localization and robustness of topological markers of TCPs}\label{sec::framework}

\subsection{I.1 Conventional momentum-space classification}

Consider a translationally invariant $d$-dimensional system described by a Hamiltonian $\hat{H}.$ Suppose the system is invariant under a spatial symmetry $G_{\mathcal{S}},$ where $\mathcal{S}$ denotes collectively the set of points that are left invariant by $G_{\mathcal{S}}$ in a system with periodic boundary conditions. If we denote by $\hat{g}_{\mathcal{S}}$ the Hilbert space representation of $G_{\mathcal{S}},$ then
\begin{equation}
    [\hat{g}_{\mathcal{S}},\hat{H}]=0.
\end{equation}
Depending on the dimensionality of the system, when the spatial symmetry acts on a general position vector $\textbf{r},$ it can leave invariant a subset of its components. We denote by $\textbf{r}_I (\textbf{r}_G)$ the components of $\textbf{r}$ that are invariant (changed) by the operation $G_{\mathcal{S}}.$ The number of components of $\textbf{r}_I (\textbf{r}_G)$ is $d_I (d_G)$ such that $d_I+d_G=d.$

We study the Bloch Hamiltonian $h(\textbf{k})$ obtained by Fourier transforming \textit{\textbf{only}} the spatial coordinates $\textbf{r}_G.$ Thus, $\textbf{k}$ is a wave vector with $d_G$ components. The momentum-space expression of $\hat{g}_{\mathcal{S}}$ is
\begin{equation}
    \hat{g}_{\mathcal{S}}=\sum_{\textbf{k}}\widetilde{g}_{\mathcal{S}}(\textbf{k})\otimes\ket{G_{\mathcal{S}}\textbf{k}}\bra{\textbf{k}},
\end{equation}
so the action of the spatial symmetry on the Bloch Hamiltonian is then
\begin{equation}
\tilde{g}_{\mathcal{S}} (\mathbf{k})h(\mathbf{k}) \tilde{g}^{-1}_{\mathcal{S}}(\mathbf{k})= h(G_{\mathcal{S}}\mathbf{k}).
\end{equation}
At the momenta $\textbf{k}^{\text{inv}}$ that are invariant under $G_{\mathcal{S}},$ we have
\begin{equation}
[\tilde{g}_{\mathcal{S}}(\mathbf{k}^{\text{inv}}),h(\mathbf{k}^{\text{inv}})]=0.
\end{equation}
Thus, the occupied states $\{\ket{u^n_{j\textbf{k}^{\text{inv}}}}\}$ can be classified according to the eigenvalues $\{e^{i\phi_j(\textbf{k}^\text{inv})}\}$ of $\tilde{g}_{\mathcal{S}}(\textbf{k}^{\text{inv}}).$ Let us denote the projector into the occupied eigenstates of a given subspace by $\mathcal{P}_{j\mathbf{k}^{\text{inv}}}=\sum_{n\in \text{occ}}\ket{u^n_{j\textbf{k}^{\text{inv}}}}\bra{u^n_{j\textbf{k}^{\text{inv}}}}$. The conventional classification of a large class of TCPs is based on determining the topology encoded by $\mathcal{P}_{j\mathbf{k}^{\text{inv}}}.$ 

The set of topological invariants $\{\tau_j(\textbf{k}^{\text{inv}})\}$ that characterize the invariant subspaces depend on their effective dimensionality $d_s$ and the other symmetries of the system. Examples of possible invariants that could be relevant include occupation numbers (e.g. \cite{Hughes2011}), winding numbers (e.g. \cite{Zhang2013}) and Chern numbers (e.g. \cite{Fu2011}). Such topological invariants $\{\tau_j(\textbf{k}^{\text{inv}})\}$ can generally be written as the average over occupied states of an operator $\mathcal{F}$ that is a function of $\mathcal{P}_{j\mathbf{k}^{\text{inv}}}:$
\begin{equation}
\tau_j(\mathbf{k}_{\text{inv}})=\text{Tr}_{\circ}\left[\mathcal{F}(\mathcal{P}_{j\mathbf{k}_{\text{inv}}})\right],
\end{equation} 
where $\text{Tr}_{\circ}$ refers to the trace over the non-spatial degrees of freedom as well as the $d_I$ spatial degrees of freedom that do not get transformed by $G_{\mathcal{S}}.$ For a broad range of TCPs, the function $\mathcal{F}$ is written as a sum of products of projection operators and position operators that are components of $\textbf{r}_{I},$ such that it projects the trace into a sum over occupied states. Examples of the form of $\mathcal{F}$ are
\begin{equation}
\mathcal{F}(\mathcal{P})\propto \left\{
\begin{array}{ll}
\mathcal{P} \hspace{2.8cm}\text{(e.g. $0D$ occupation number \cite{Hughes2011} for $1D$ TCP),} \\
\mathcal{P}\left[\hat{X},\mathcal{P}\right]\hspace{1.85cm}\text{(e.g. $1D$ winding number \cite{MS2014} for $2D$ TCP),} \\
\mathcal{P}\left[\left[\hat{X},\mathcal{P}\right], \left[\hat{Y},\mathcal{P}\right]\right] \hspace{0.3cm}\text{(e.g. $2D$ Chern number \cite{Prodan2011} for $3D$ TCP),}
\end{array} \right.\label{Eq_examples}
\end{equation}
where $\hat{X},\hat{Y}$ are the $x,y$ components of the position operator.  These commutators are the basis-independent generalizations of momentum-space derivatives that naturally arise en the calculations of Berry connections. 

\subsection{I.2 Basis-independent topological invariants of TCPs}

We now set out to obtain topological invariants with localized topological markers using the momentum-space invariants $\{\tau_j(\textbf{k}^{\text{inv}})\}.$ Let us now define the quantities
\begin{equation}
    T^{g}_{\mathcal{S}}\equiv\sum_{j,\textbf{k}^{\text{inv}}}  e^{i\phi_j(\textbf{k}^{\text{inv}})}  \tau_j(\textbf{k}^{\text{inv}}).
\end{equation} 
Since the $\{T^{g}_{\mathcal{S}}\}$ are linear combinations of the momentum-space topological invariants, they are themselves topological invariants of the TCP. We will now show that these topological invariants have topological markers that are that are exponentially localized and are robust to the presence of impurities in the system. This will be done by expressing them in terms of the projector in the full ground state given by
\begin{equation}
    \mathcal{P}_G=\sum_{j\textbf{k}}\mathcal{P}_{\textbf{k}j}.
\end{equation}
We can rewrite them in a basis-independent form:
\begin{eqnarray}
    T^{g}_{\mathcal{S}}&=&\sum_{j,\textbf{k}^{\text{inv}}}  e^{i\phi_j(\textbf{k}^{\text{inv}})} \tau_j(\textbf{k}^{\text{inv}})\\
    &=&\sum_{j,\textbf{k}^{\text{inv}}}   \text{Tr}_{\circ}\left[e^{i\phi_j(\textbf{k}^{\text{inv}})}\mathcal{F}(\mathcal{P}_{j,\mathbf{k}^{\text{inv}}})\right]\\
    &=&\sum_{j,\textbf{k}^{\text{inv}}}   \text{Tr}_{\circ}\left[e^{i\phi_j(\textbf{k}^{\text{inv}})}\mathcal{P}_{j\mathbf{k}_{\text{inv}}}\mathcal{F}(\mathcal{P}_{j,\mathbf{k}^{\text{inv}}})\right]\\
    &=&\sum_{j,\textbf{k}^{\text{inv}}}  \text{Tr}_{\circ}\left[\mathcal{P}_{j\mathbf{k}^{\text{inv}}}e^{i\phi_j(\textbf{k}^{\text{inv}})}\mathcal{P}_{j\mathbf{k}^{\text{inv}}}\mathcal{F}(\mathcal{P}_{j,\mathbf{k}^{\text{inv}}})\right].
\end{eqnarray}
Now, since $\hat{\textbf{r}}_I$ and $\mathcal{P}_{j\textbf{k}^{\text{inv}}}$ commute with $\widetilde{g}_{\mathcal{S}}(\textbf{k}^{\text{inv}}),$ then $[\widetilde{g}_{\mathcal{S}}(\textbf{k}^{\text{inv}}),\mathcal{F}(\mathcal{P}_{j\textbf{k}^{\text{inv}}})]=0.$  This implies that $\mathcal{F}(\mathcal{P}_{j\textbf{k}^{\text{inv}}})$ is block-diagonal in the basis of $\widetilde{g}_{\mathcal{S}}(\textbf{k}^{\text{inv}}),$ and we can thus write $\sum_j\mathcal{F}(\mathcal{P}_{j\mathbf{k}^{\text{inv}}})=\mathcal{F}\left(\sum_j\mathcal{P}_{j\mathbf{k}^{\text{inv}}}\right).$ Furthermore, by assumption, the terms that make up $\mathcal{F}(\mathcal{P}_G)$ are invariant under translations along the components of $\hat{\textbf{r}}_G.$ This means that $\mathcal{F}$ is block-diagonal with respect to $\mathbf{k}.$ As a consequence
\begin{eqnarray}
    \sum_{\textbf{k}}\mathcal{F}\left(\mathcal{P}_{\mathbf{k}}\right)\otimes \ket{\textbf{k}}\bra{\textbf{k}}&=&\mathcal{F}\left(\sum_{\textbf{k}}\mathcal{P}_{\mathbf{k}}\otimes \ket{\textbf{k}}\bra{\textbf{k}}\right)=\mathcal{F}(\mathcal{P}_G).
\end{eqnarray} 
Using these properties of $\mathcal{F},$ we then obtain
\begin{eqnarray}
    T^{g}_{\mathcal{S}}&=&\sum_{\textbf{k}^{\text{inv}}}  \text{Tr}_{\circ}\left[\left(\sum_{j_1}\mathcal{P}_{j_1\mathbf{k}^{\text{inv}}}\right)\widetilde{g}_{\mathcal{S}}(\textbf{k}^{\text{inv}})\left(\sum_{j_2}\mathcal{P}_{j_2\mathbf{k}^{\text{inv}}}\right)\mathcal{F}\left(\sum_{j_3}\mathcal{P}_{j_3\mathbf{k}^{\text{inv}}}\right)\right]\\
    &=&\sum_{\textbf{k}^{\text{inv}}}  \text{Tr}_{\circ}\left[\left(\mathcal{P}_{\mathbf{k}^{\text{inv}}}\right)\widetilde{g}_{\mathcal{S}}(\textbf{k}^{\text{inv}})\left(\mathcal{P}_{\mathbf{k}^{\text{inv}}}\right)\left(\mathcal{F}(\mathcal{P}_{\mathbf{k}^{\text{inv}}})\right)\right]\\
    &=&\sum_{\textbf{k}^{\text{inv}}}   \bra{\textbf{k}^{\text{inv}}}\text{Tr}_{\circ}\left[\left(\mathcal{P}_{\textbf{k}^{\text{inv}}}\otimes\ket{\textbf{k}^{\text{inv}}}\bra{\textbf{k}^{\text{inv}}}\right)\left(\tilde{g}_{\mathcal{S}}(\textbf{k}^{\text{inv}})\otimes\ket{\textbf{k}^{\text{inv}}}\bra{\textbf{k}^{\text{inv}}}\right)\right. \nonumber\\
    &&\hspace{1.5cm}\times \left. \left(\mathcal{P}_{\textbf{k}^{\text{inv}}}\otimes\ket{\textbf{k}^{\text{inv}}}\bra{\textbf{k}^{\text{inv}}}\right)\left(\mathcal{F}(\mathcal{P}_{\textbf{k}^{\text{inv}}})\otimes\ket{\textbf{k}^{\text{inv}}}\bra{\textbf{k}^{\text{inv}}}\right)\right]\ket{\textbf{k}^{\text{inv}}}\\
    &=&\sum_{\textbf{k}^{\text{inv}}}   \bra{\textbf{k}^{\text{inv}}}\text{Tr}_{\circ}\left[\left(\sum_{\textbf{q}_1}\mathcal{P}_{\mathbf{q}_1}\otimes\ket{\textbf{q}_1}\bra{\textbf{q}_1}\right)\left(\sum_{\textbf{q}_2}\tilde{g}_{\mathcal{S}}(\textbf{q}_2)\otimes\ket{G_{\mathcal{S}}\textbf{q}_2}\bra{\textbf{q}_2}\right)\right. \nonumber\\
    &&\hspace{1.5cm}\times \left. \left(\sum_{\textbf{q}_3}\mathcal{P}_{\mathbf{q}_3}\otimes\ket{\textbf{q}_3}\bra{\textbf{q}_3}\right)\left(\sum_{\textbf{q}_4}\mathcal{F}(\mathcal{P}_{\mathbf{q}_4})\otimes\ket{\textbf{q}_4}\bra{\textbf{q}_4}\right)\right]\ket{\textbf{k}^{\text{inv}}}\nonumber\\
    &=&\sum_{\textbf{k}}   \bra{\textbf{k}}\text{Tr}_{\circ}\left[\left(\sum_{\textbf{q}_1}\mathcal{P}_{\mathbf{q}_1}\otimes\ket{\textbf{q}_1}\bra{\textbf{q}_1}\right)\left(\sum_{\textbf{q}_2}\tilde{g}_{\mathcal{S}}(\textbf{q}_2)\otimes\ket{G_{\mathcal{S}}\textbf{q}_2}\bra{\textbf{q}_2}\right)\right. \nonumber\\
    &&\hspace{2cm}\times \left.\left(\sum_{\textbf{q}_3}\mathcal{P}_{\mathbf{q}_3}\otimes\ket{\textbf{q}_3}\bra{\textbf{q}_3}\right)\mathcal{F}\left(\sum_{\textbf{q}_4}\mathcal{P}_{\mathbf{q}_4}\otimes\ket{\textbf{q}_4}\bra{\textbf{q}_4}\right)\right]\ket{\textbf{k}}\\
    &=&\sum_{\textbf{k}}   \bra{\textbf{k}}\text{Tr}_{\circ}\left[\mathcal{P}_G\hat{g}_{\mathcal{S}}\mathcal{P}_G\mathcal{F}(\mathcal{P}_G)\right]\ket{\textbf{k}}\\
    &=&\text{Tr}[\overline{g}_{\mathcal{S}} \mathcal{F}(\mathcal{P}_{\text{G}})]. \label{Eq_gen}
\end{eqnarray}
which is the basis-independent expression quoted in the main text for these topological invariants.

\subsection{I.3 Localized topological markers}

We now show that the presence of the projected symmetry operator $\overline{g}_{\mathcal{S}}$ in our expression for $T^{g}_{\mathcal{S}}$ implies that the topological markers are spatially localized. Consider a position $\mathbf{r}_G$ far away from the fixed points $\mathcal{S}$ (a condition we denote by $\vert \mathcal{S}-\mathbf{r}_G\vert\gg \zeta$), which implies that $\vert G_{\mathcal{S}}\mathbf{r}_G-\mathbf{r}_G\vert\gg \zeta.$ We know that for gapped systems, the projection operator has the long-distance behavior
\begin{equation}
    \vert\bra{\textbf{r}_{G_1} ,\alpha}\mathcal{P}_G\ket{\textbf{r}_{G_2},\beta}\vert<\mathcal{O}\left(e^{-\vert \textbf{r}_{G_1}-\textbf{r}_{G_2}\vert/\zeta}\right)\quad \text{when} \quad \zeta \ll \vert \textbf{r}_{G_1}-\textbf{r}_{G_2}\vert,
\end{equation}
where $\alpha,\beta$ denote all internal and remaining spatial degrees of freedom.
Furthermore, since $\mathcal{F}$ is only dependent on $\mathcal{P}_G$ and the components of $\hat{\textbf{r}_I},$ then
\begin{equation}
\vert\bra{\textbf{r}_{G_1} ,\alpha}\mathcal{F}(\mathcal{P}_G)\ket{\textbf{r}_{G_2},\beta}\vert<\mathcal{O}\left(e^{-\vert \textbf{r}_{G_1}-\textbf{r}_{G_2}\vert/\zeta}\right)\quad \text{when} \quad \zeta \ll \vert \textbf{r}_{G_1}-\textbf{r}_{G_2}\vert.
\end{equation}
Using this, we can then write the topological marker $\mathcal{T}^{g}_{\mathcal{S}}(\textbf{r}_G)=\bra{\textbf{r}_G}\text{Tr}_{\circ}\left[\overline{g}_{\mathcal{S}}\mathcal{F}(\mathcal{P}_G)\right]\ket{\textbf{r}_G}$ as
\begin{eqnarray}
    \mathcal{T}^{g}_{\mathcal{S}}(\textbf{r}_G)&=&\text{Tr}_{\circ}[\bra{\textbf{r}'_G}\hat{g}_{\mathcal{S}} \mathcal{P}_G\mathcal{F}(\mathcal{P}_{\text{G}})\ket{\textbf{r}_G}]\\
    &=&\sum_{\textbf{r}_{G1}\textbf{r}_{G2}}\text{Tr}_{\circ}\left[\bra{\textbf{r}_G}\hat{g}_{\mathcal{S}}\ket{\textbf{r}_{G1}}\bra{\textbf{r}_{G1}} \mathcal{P}_G\ket{\textbf{r}_{G2}}\bra{\textbf{r}_{G2}}\mathcal{F}(\mathcal{P}_{\text{G}})\ket{\textbf{r}_G}\right]\\
    &=&\sum_{\textbf{r}_{G1}\textbf{r}_{G2}}\text{Tr}_{\circ}\left[\delta_{\textbf{r}_G,G_{\mathcal{S}}\textbf{r}_{G1}}\bra{\textbf{r}_{G1}} \mathcal{P}_G\ket{\textbf{r}_{G2}}\bra{\textbf{r}_{G2}}\mathcal{F}(\mathcal{P}_{\text{G}})\ket{\textbf{r}_G}\right]\\
    &=&\sum_{\textbf{r}_{G2}}\text{Tr}_{\circ}\left[\bra{G^{-1}_{\mathcal{S}}\textbf{r}_G} \mathcal{P}_G\ket{\textbf{r}_{G2}}\bra{\textbf{r}_{G2}}\mathcal{F}(\mathcal{P}_{\text{G}})\ket{\textbf{r}_G}\right]\\
    &=&\sum_{\vert \textbf{r}_{G2}-\textbf{r}_G\vert<\zeta}\text{Tr}_{\circ}\left[\bra{G^{-1}_{\mathcal{S}}\textbf{r}_G} \mathcal{P}_G\ket{\textbf{r}_{G2}}\bra{\textbf{r}_{G2}}\mathcal{F}(\mathcal{P}_{\text{G}})\ket{\textbf{r}_G}\right]\nonumber\\
    &&+\sum_{\vert \textbf{r}_{G2}-G^{-1}_{\mathcal{S}}\textbf{r}_G\vert<\zeta}\text{Tr}_{\circ}\left[\bra{G^{-1}_{\mathcal{S}}\textbf{r}_G} \mathcal{P}_G\ket{\textbf{r}_{G2}}\bra{\textbf{r}_{G2}}\mathcal{F}(\mathcal{P}_{\text{G}})\ket{\textbf{r}_G}\right]\\
     &&+\sum_{\substack{\vert \textbf{r}_{G2}-G^{-1}_{\mathcal{S}}\textbf{r}_G\vert>\zeta\\
     \vert \textbf{r}_{G2}-\textbf{r}_G\vert>\zeta}}\text{Tr}_{\circ}\left[\bra{G^{-1}_{\mathcal{S}}\textbf{r}_G} \mathcal{P}_G\ket{\textbf{r}_{G2}}\bra{\textbf{r}_{G2}}\mathcal{F}(\mathcal{P}_{\text{G}})\ket{\textbf{r}_G}\right].
\end{eqnarray}
Now, we have that 
\begin{eqnarray}
    A_1&=&\left\vert\sum_{\vert \textbf{r}_{G2}-\textbf{r}_G\vert<\zeta}\text{Tr}\left[\bra{G^{-1}_{\mathcal{S}}\textbf{r}_G} \mathcal{P}_G\ket{\textbf{r}_{G2}}\bra{\textbf{r}_{G2}}\mathcal{F}(\mathcal{P}_{\text{G}})\ket{\textbf{r}_G}\right]\right\vert<\mathcal{O}\left(e^{-\vert G^{-1}_{\mathcal{S}}\textbf{r}_G-\textbf{r}_{G}\vert/\zeta}\right)\\
    A_2&=&\left\vert\sum_{\vert \textbf{r}_{G2}-G^{-1}_{\mathcal{S}}\textbf{r}_G\vert<\zeta}\text{Tr}\left[\bra{G^{-1}_{\mathcal{S}}\textbf{r}_G} \mathcal{P}_G\ket{\textbf{r}_{G2}}\bra{\textbf{r}_{G2}}\mathcal{F}(\mathcal{P}_{\text{G}})\ket{\textbf{r}_G}\right]\right\vert<\mathcal{O}\left(e^{-\vert G^{-1}_{\mathcal{S}}\textbf{r}_G-\textbf{r}_{G}\vert/\zeta}\right)\\
    A_3&=&\left\vert\sum_{\substack{\vert \textbf{r}_{G2}-G^{-1}_{\mathcal{S}}\textbf{r}_G\vert>\zeta\\
     \vert \textbf{r}_{G2}-\textbf{r}_G\vert>\zeta}}\text{Tr}\left[\bra{G^{-1}_{\mathcal{S}}\textbf{r}_G} \mathcal{P}_G\ket{\textbf{r}_{G2}}\bra{\textbf{r}_{G2}}\mathcal{F}(\mathcal{P}_{\text{G}})\ket{\textbf{r}_G}\right]\right\vert \ll A_{1,2}.
\end{eqnarray}
We thus conclude that
\begin{eqnarray}
    \vert\mathcal{T}^{g}_{\mathcal{S}}(\textbf{r}_G)\vert &<&\mathcal{O}\left(e^{-\vert G_{\mathcal{S}}\textbf{r}_G-\textbf{r}_{G}\vert/\zeta}\right) \quad \text{when} \quad \zeta \ll \vert G_{\mathcal{S}}\textbf{r}_G-\textbf{r}_{G}\vert.
\end{eqnarray}
which implies that the topological marker is concentrated around the fixed points of the spatial symmetry.

\subsection{I.4 Robustness to impurities placed far from fixed points.}
We now show that although the conventional momentum-space invariants cannot be used in the presence of impurities, the invariants $T^{g}_{\mathcal{S}}$ can remain robustly quantized. Suppose we add an impurity $V_{\text{imp}}(\textbf{r})=V_0\delta_{\textbf{r},\textbf{R}_0}$ at a position $\textbf{R}_0.$ 
The projection operator of the new ground state can be written as $\mathcal{P}'_G=\mathcal{P}_G+\delta \mathcal{P}_G.$ Since the system is gapped, $\delta \mathcal{P}_G$  only has support in an exponentially small neighborhood of size $\zeta$ centered at the impurity:
\begin{eqnarray}
\vert\bra{\textbf{r}_G,\alpha}\delta\mathcal{P}_G\ket{\textbf{R}_0,\beta}\vert< \mathcal{O}\left(e^{-\left(\vert \textbf{r}_G-\textbf{R}_0\vert\right)/\zeta }\right).
\end{eqnarray}
Let us now define a function $\Gamma(\mathcal{P}'_G)$ by
\begin{equation}
    \delta\Gamma(\mathcal{P}'_G)\equiv\mathcal{P}'_G\mathcal{F}(\mathcal{P}'_G)-\mathcal{P}_G\mathcal{F}(\mathcal{P}_G).
\end{equation}
The function $\Gamma(\mathcal{P}'_G)$ must vanish when $\delta \mathcal{P}_G=0,$ so it must be a sum of terms all of which possess at least one factor of $\delta \mathcal{P}_G.$ The long-distanceIt follows that 
\begin{eqnarray}
\vert \bra{\textbf{r}_G,\alpha}\Gamma(\mathcal{P}'_G) \ket{\textbf{R}_0,\beta}\vert<\mathcal{O}\left(e^{-\vert \textbf{r}_G-\textbf{R}_0\vert/\zeta }\right) \quad \text{for} \quad \vert \textbf{r}_G-\textbf{R}_0\vert\gg\zeta
\end{eqnarray}
For an impurity placed far away from the points in $\mathcal{S},$ the topological invariant is modified as
\begin{eqnarray}
    T^{g'}_{\mathcal{S}}(\textbf{R}_0)-T^{g}_{\mathcal{S}}&=&\text{Tr}\left[\hat{g}_{\mathcal{S}}\Gamma(\mathcal{P}'_G)\right]\\
    &=&\sum_{\textbf{r}_G\textbf{r}_{G1}}\text{Tr}_{\circ}\left[\bra{\textbf{r}_G}\hat{g}_{\mathcal{S}}\ket{\textbf{r}_{G1}}\bra{\textbf{r}_{G1}}\Gamma(\mathcal{P}'_G)\ket{\textbf{r}_G}\right]\\
    &=&\sum_{\textbf{r}_G\textbf{r}_{G1}}\text{Tr}_{\circ}\left[\delta_{\textbf{r}_G,G_{\mathcal{S}}\textbf{r}_{G1}}\bra{\textbf{r}_{G1}}\Gamma(\mathcal{P}'_G)\ket{\textbf{r}_G}\right]\\
    &=&\sum_{\textbf{r}_G}\text{Tr}_{\circ}\left[\bra{G^{-1}_{\mathcal{S}}\textbf{r}_G}\Gamma(\mathcal{P}'_G)\ket{\textbf{r}_G}\right]\\
    &=&\sum_{\vert G^{-1}_\mathcal{S}\textbf{r}_{G}-\textbf{R}_0\vert<\zeta}\text{Tr}_{\circ}\left[\bra{G^{-1}_{\mathcal{S}}\textbf{r}_G} \Gamma(\mathcal{P}'_G)\ket{\textbf{r}_G}\right]\nonumber\\
    &&+\sum_{\vert \textbf{r}_{G}-\textbf{R}_0\vert<\zeta}\text{Tr}_{\circ}\left[\bra{G^{-1}_{\mathcal{S}}\textbf{r}_G} \Gamma(\mathcal{P}'_G)\ket{\textbf{r}_G}\right]\nonumber\\
     &&+\sum_{\substack{\vert G^{-1}_{\mathcal{S}}\textbf{r}_{G}-\textbf{R}_0\vert>\zeta\\
     \vert \textbf{r}_{G}-\textbf{R}_0\vert>\zeta}}\text{Tr}_{\circ}\left[\bra{G^{-1}_{\mathcal{S}}\textbf{r}_G} \Gamma(\mathcal{P}'_G)\ket{\textbf{r}_G}\right].
\end{eqnarray}
Now, we have that 
\begin{eqnarray}
    B_1&=&\left\vert\sum_{\vert G^{-1}_\mathcal{S}\textbf{r}_{G}-\textbf{R}_0\vert<\zeta}\text{Tr}_{\circ}\left[\bra{G^{-1}_{\mathcal{S}}\textbf{r}_G} \Gamma(\mathcal{P}'_G)\ket{\textbf{r}_G}\right]\right\vert<\mathcal{O}\left(e^{-\vert \textbf{R}_0-G_{\mathcal{S}}\textbf{R}_{0}\vert/\zeta}\right)\\
    B_2&=&\left\vert\sum_{\vert \textbf{r}_{G}-\textbf{R}_0\vert<\zeta}\text{Tr}_{\circ}\left[\bra{G^{-1}_{\mathcal{S}}\textbf{r}_G} \Gamma(\mathcal{P}'_G)\ket{\textbf{r}_G}\right]\right\vert<\mathcal{O}\left(e^{-\vert G^{-1}_{\mathcal{S}}\textbf{R}_0-\textbf{R}_{0}\vert/\zeta}\right)\\
    B_3&=&\left\vert\sum_{\substack{\vert G^{-1}_{\mathcal{S}}\textbf{r}_{G}-\textbf{R}_0\vert>\zeta\\
     \vert \textbf{r}_{G}-\textbf{R}_0\vert>\zeta}}\text{Tr}_{\circ}\left[\bra{G^{-1}_{\mathcal{S}}\textbf{r}_G} \Gamma(\mathcal{P}'_G)\ket{\textbf{r}_G}\right]\right\vert \ll B_{1,2}.
\end{eqnarray}
We thus conclude that
\begin{equation}
   \vert T^{g'}_{\mathcal{S}}(\textbf{R}_0)-T^{g}_{\mathcal{S}}\vert<\mathcal{O}\left( e^{-\vert G_{\mathcal{S}}\textbf{R}_0- \textbf{R}_0\vert/\zeta}\right),
\end{equation}
which implies that impurities in the system will not affect the quantization of the invariants $T^{g}_{\mathcal{S}}$ as long their distance to $\mathcal{S}$ is many times bigger than a correlation length $\zeta.$

%%%%%%%%%%%%%%%%%%%%%%%%%%%%%%%%%%%%%%%%%%%%%%%

\section{Section II:  Inversion-symmetric topological insulators in one dimension}

\subsection{II.1 Conventional classification}

Consider topological insulators in one-dimensional protected by inversion symmetry. With periodic boundary conditions, an inversion symmetry operation $P_{\mathcal{S}}$ leaves invariant two positions $\mathcal{S}=\{R+r_0,R+1+L_x/2+r_0\}$ on the lattice with $R\leq L_x/2.$ The number $r_0$ is a fixed position within the unit cell that can take the values $0$ or $1/2,$ where we set the unit cell length to be $a=1.$ Consider the momentum-space expansion of the inversion operator
\begin{equation}
    \hat{p}_{\mathcal{S}}=\sum_{k_x}\widetilde{p}_{\mathcal{S}}(k_x)\otimes\ket{P_{\mathcal{S}}k_x}\bra{k_x}.
\end{equation}
The conventional momentum-space classification is based on counting the number of occupied states $n^{\pm}_{\mathcal{S}}(k^{\text{inv}}_x)$ with eigenvalues of the operator $\widetilde{p}_{\mathcal{S}}(k_x)$ at the invariant momenta $k^{\text{inv}}_x=0,\pi.$  The invariant is given by \cite{Fang2013b}
\begin{eqnarray}
\Delta^p_{\mathcal{S}}&=&\sum_{k_x=0,\pi}\left[n_{\mathcal{S}}^{(+)}(k_x)-n_{\mathcal{S}}^{(-)}(k_x)\right].\nonumber
\end{eqnarray}
which is an integer invariant.

\subsection{II.2 Basis-independent classification}

We now discuss the topological invariants $T^p_{\mathcal{S}}$ for this TCP. The topological invariants at the invariant momenta are 
\begin{equation}
    \tau_{\pm}(k^{\text{inv}}_x)=n^{\pm}_{\mathcal{S}}(k^{\text{inv}}_x)=\text{Tr}_\circ\left[\mathcal{P}_{\pm,k^{\text{inv}}_x}\right].
\end{equation} 
It is clear then that, in this case, $\mathcal{F}(\mathcal{P}_{\pm,k^{\text{inv}}_x})=\mathcal{P}_{\pm,k^{\text{inv}}_x}.$ By using these invariants in Eq.\ref{Eq_gen}, we find
\begin{eqnarray}
T^p_{\mathcal{S}}&=&\sum_{k_x=0,\pi}\left[(+1)n_{\mathcal{S}}^{(+)}(k_x)+(-1)n_{\mathcal{S}}^{(-)}(k_x)\right]=\text{Tr}\left[\overline{p}_{\mathcal{S}} \cdot\mathcal{P}_{\pm,k^{\text{inv}}_x}\right]=\text{Tr}\left[\overline{p}_{\mathcal{S}}\right].\nonumber
\end{eqnarray}
In this particular example, the linear combination $T^p_{\mathcal{S}}$ of momentum-space topological invariants is precisely the topological invariant $\Delta^p_{\mathcal{S}}$ that was proposed for these TCPs \cite{Fang2013b}.
The topological marker for this invariant satisfies
\begin{eqnarray}
    \vert\mathcal{T}^{p}_{\mathcal{S}}(x)\vert &<&\mathcal{O}\left(e^{-2\vert x-\mathcal{S}\vert/\zeta}\right) \quad \text{when} \quad \zeta \ll \vert x-\mathcal{S}\vert,
\end{eqnarray}
and an impurity placed at $x=X_0$ will change an invariant centered at $\mathcal{S}$ by an exponentially small amount
\begin{equation}
   \vert T^{p'}_{\mathcal{S}}(X_0)-T^{p}_{\mathcal{S}}\vert<\mathcal{O}\left( e^{-2\vert X_0- \mathcal{S}\vert/\zeta}\right) \quad \text{when} \quad \zeta \ll \vert X_0-\mathcal{S}\vert,
\end{equation}
where the notation $\vert X_0 -\mathcal{S} \vert$ means the shortest distance to any of the invariant points in the set $\mathcal{S}.$

\subsection{II.3 Model example}

The model presented in the main text is
\begin{equation}
    h(k_x)=\sin k_x \sigma_2+(m-\cos k_x)\sigma_1,
\end{equation}
where the Pauli matrices $\sigma_a$ act on two orbitals $\{A,B\}.$ The momentum-space inversion operator is $\widetilde{p}_{\mathcal{S}}(k_{x}^{\text{inv}})=e^{-2i k_{x}^{\text{inv}} r_0} \sigma_1.$ The topological phase occurs for $\vert m\vert<1.$ For the figures in the main text we chose $m=0.3.$

%%%%%%%%%%%%%%%%%%%%%%%%%%%%%%%%%%%%%%%%%%%%%%%

\begin{figure}
\centering
\includegraphics[scale=0.19]{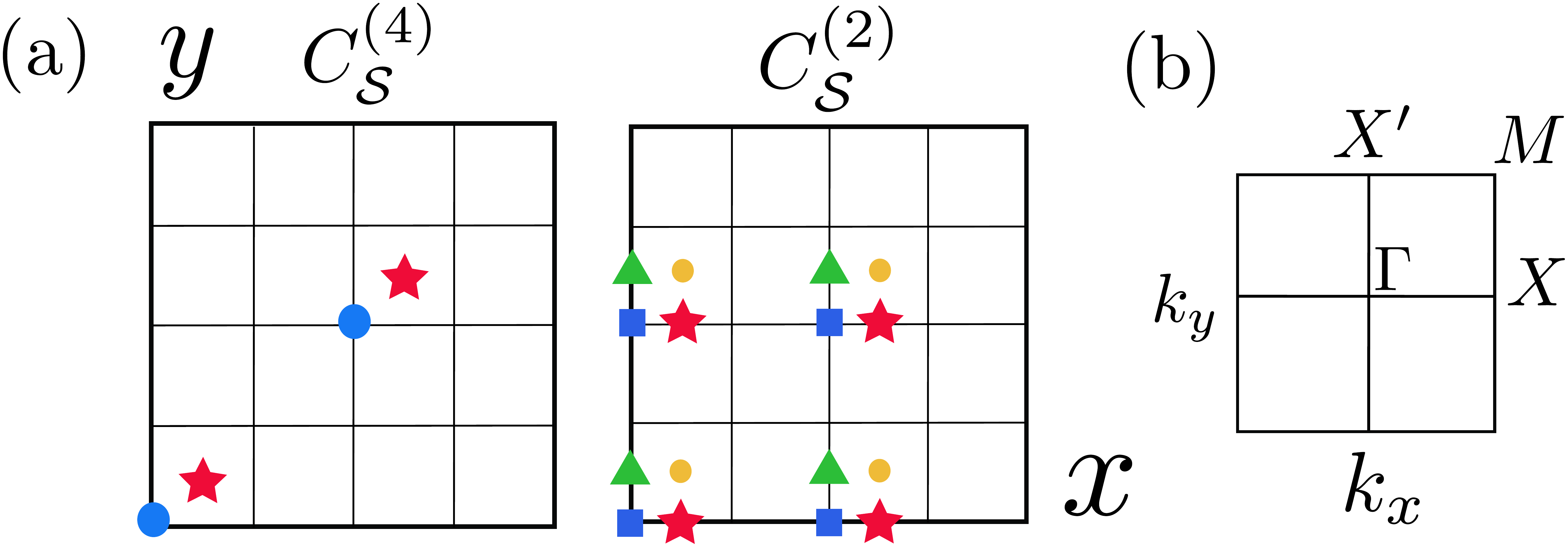}
\caption{\textbf{(a)} Invariant points under $C^{(4)}$ (left) and $C^{(2)}$ (right). For example, in the former case, red stars map to themselves under four-fold rotations centered at any of the stars, and so-forth for the remaining points. \textbf{(b)} Invariant momenta in the BZ.}\label{Fig_rot_S}
\end{figure}

\section{Section III: Rotationally-symmetric topological superconductors in two dimensions}

\subsection{III.1 Conventional classification}

Consider topological superconductors in the absence of time-reversal symmetry and protected by rotational symmetries. With periodic boundary conditions, a given $n$-fold rotation operation $C^{(n)}_{\mathcal{S}}$ leaves invariant a set $\mathcal{S}=\{\textbf{R}_i+\mathbf{r}_0\}$ of positions on the lattice, as illustrated in Fig. \ref{Fig_rot_S}.  In particular, we will consider models that satisfy $[\hat{c}^{(n=2,4)}_{\mathcal{S}},\hat{H}]=0$, where $\hat{c}^{(n)}_{\mathcal{S}}$ is the Hilbert space representation of $C^{(n)}_{\mathcal{S}}.$ By denoting the primitive vectors as $2\textbf{a}_{1}$  and $2\textbf{a}_{2},$ then four-fold rotations allow the positions $\mathbf{r}_0\in \{\textbf{0}, \textbf{a}_1+\textbf{a}_2\},$ whereas two-fold rotations allow $\textbf{r}_0\in\{\textbf{0},\textbf{a}_1,\textbf{a}_2, \textbf{a}_1+\textbf{a}_2\}.$ Consider the momentum-space expansion of the rotation operators
\begin{equation}
    \hat{c}^{(n)}_{\mathcal{S}}=\sum_{\textbf{k}}\widetilde{c}^{(n)}_{\mathcal{S}}(\textbf{k})\otimes\ket{C^{(n)}_{\mathcal{S}}\textbf{k}}\bra{\textbf{k}}.
\end{equation}
The conventional momentum-space classification is based on counting eigenvalues of the specific operator $\widetilde{c}^{(n)}_{\mathcal{S}}(\textbf{k})\vert_{\textbf{r}_0=\textbf{0}}.$ For a given invariant momentum $\mathbf{\Pi}^{(n)} (=\Gamma,X,M_{i}),$ the possible eigenvalues of $\widetilde{c}^{(n)}_{\mathcal{S}}(\textbf{k})\vert_{\textbf{r}_0=\textbf{0}}$ are $\Pi^{(n)}_p=e^{i\pi(2p-1)/n},$ for $p=1,2,\ldots, n.$ Let us denote by $\#\Pi^{(n)}_p$ the number of occupied states with eigenvalue $\Pi^{(n)}_p$ at the momentum $\mathbf{\Pi}^{(n)}.$ The topological invariants are then obtained by defining a relative occupation number
\begin{equation}
    [\Pi^{(n)}_p]=\#\Pi^{(n)}_p-\#\Gamma^{(n)}_p.
\end{equation}
In particular, as we mentioned in the main text, the topological invariants for systems that are symmetric under four-fold rotations are $([X],[M_1],[M_2]).$

\subsection{III.2 Basis-independent classification}

We now proceed to adapt Eq.\ref{Eq_gen} to these TCPs. We define the topological invariants of the invariant subspaces as the occupation numbers $n^{(n,p)}_{\mathcal{S}}(\mathbf{\Pi}^{(n)})$ of the eigenvalues of the operator $\widetilde{c}^{(n)}_{\mathcal{S}}(\textbf{k}):$ 
\begin{equation}
\tau_{np}(\mathbf{\Pi}^{(n)})=n^{(n,p)}_{\mathcal{S}}(\mathbf{\Pi}^{(n)})=\text{Tr}_{\circ}\left[\mathcal{P}_{j\textbf{k}^{\text{inv}}}\right],    
\end{equation}
where we again have used $\mathcal{F}(\mathcal{P}_{j\textbf{k}^{\text{inv}}})=\mathcal{P}_{j\textbf{k}^{\text{inv}}}.$  From Eq.\ref{Eq_gen}, we know that this implies that
\begin{equation}
T^{c(n)}_{\mathcal{S}}=\text{Tr}\left[\overline{c}^{(n)}_{\mathcal{S}}\mathcal{F}(\mathcal{P}_G)\right]=\text{Tr}\left[\overline{c}^{(n)}_{\mathcal{S}}\right].   \label{Eq_rot} 
\end{equation}
In contrast with the inversion-symmetric case, the conventional momentum-space topological invariants are not exactly  the invariants $T^{c(n)}_{\mathcal{S}}.$ We can, nevertheless, find a relation between both. The rotation operator at an invariant momentum for a given set of fixed points $\mathcal{S}$ can be written in the form 
\begin{equation}
    \widetilde{c}^{(n)}_{\mathcal{S}}(\textbf{k})=\left[\widetilde{c}^{(n)}_{\mathcal{S}}(\textbf{k})\vert_{\textbf{r}_0=\textbf{0}} \right] e^{i\left(C^{(n)}_{\mathcal{S}} \mathbf{\Pi^{(n)}}-\mathbf{\Pi^{(n)}}\right)\cdot \mathbf{r}_0},
\end{equation}
which implies that their eigenvalues are
\begin{equation}
e^{i\phi_{np}(\mathbf{\Pi}^{(n)})}=e^{i\left(\pi(2p-1)/n+(C^{(n)}_{\mathcal{S}} \mathbf{\Pi}^{(n)}-\mathbf{\Pi}^{(n)})\cdot \mathbf{r}_0\right)}.  
\end{equation}
Since the $\mathbf{\Pi}^{(n)}$ are invariant momenta under rotations by definition, $(C^{(n)}_{\mathcal{S}} \mathbf{\Pi}^{(n)}-\mathbf{\Pi}^{(n)})\cdot \mathbf{r}_0=\sum_{i} m_i \textbf{b}_i$ where the $m_{1,2}$ are integers and the $\textbf{b}_{1,2}$ are primitive vectors in reciprocal space that satisfy $\textbf{b}_i\cdot (2\textbf{a}_j)=2\pi \delta_{ij}.$ With these definitions, the traces of the projected rotation operators for $n=4$ can be written as
\begin{eqnarray}
T^{c(4)}_{\mathcal{S}}&=&\sum^4_{p=1}\left(e^{i\phi_{4p}(\mathbf{\Gamma})}n^{(4,p)}_{\mathcal{S}}(\mathbf{\Gamma})+e^{i\phi_{4p}(\mathbf{M})} n^{(4,p)}_{\mathcal{S}}(\mathbf{M})\right)\\
&=&\sum^4_{p=1}\left( e^{i \pi(2p-1)/4}\#\Gamma_p+e^{i \pi(2p-1)/4-i \textbf{b}_1\cdot\textbf{r}_0}\#M_p\right)\\
&=&\sum^4_{p=1}e^{i \pi(2p-1)/4}\left( \left(1+e^{-i \textbf{b}_1\cdot\textbf{r}_0}\right)\#\Gamma_p+e^{-i \textbf{b}_1\cdot\textbf{r}_0}[M_p]\right)\nonumber\\
&=&\left\{\left(1+e^{-i \textbf{b}_1\cdot\textbf{r}_0}\right) \left(\sum_{p}e^{i \pi(2p-1)/4} \#\Gamma_p\right)+i\sqrt{2}e^{-i \textbf{b}_1\cdot\textbf{r}_0}\left([M_1]+[M_2]\right) \right\},
\end{eqnarray}
where we have used that $[M_1]=-[M_4]$ and $[M_2]=-[M_3]$ due to particle-hole symmetry. For the $n=2$ case, the topological invariants are
\begin{eqnarray}
T^{c(2)}_{\mathcal{S}}&=&\sum_{p}\left(e^{i\phi_{2p}(\mathbf{\Gamma})}n^{(2,p)}_{\mathcal{S}}(\mathbf{\Gamma})+e^{i\phi_{2p}(\mathbf{M})} n^{(2,p)}_{\mathcal{S}}(\mathbf{M})\right)\\
&=&i\left( (\#\Gamma_{1}+\#\Gamma_{3})+e^{-i (\textbf{b}_1+\textbf{b}_2)\cdot\textbf{r}_0} (\# M_1+\# M_3)+\left(e^{-i \textbf{b}_1\cdot\textbf{r}_0}+e^{-i \textbf{b}_2\cdot\textbf{r}_0}\right) \# X_1\right)\nonumber\\
&&-i\left( (\#\Gamma_{2}+\#\Gamma_{4})+e^{-i (\textbf{b}_1+\textbf{b}_2)\cdot\textbf{r}_0} (\# M_2+\# M_4)+\left(e^{-i \textbf{b}_1\cdot\textbf{r}_0}+e^{-i \textbf{b}_2\cdot\textbf{r}_0}\right) \# X_2\right)\nonumber\\
&=&i \left(1+e^{-i \textbf{b}_1\cdot\textbf{r}_0}+e^{-i \textbf{b}_2\cdot\textbf{r}_0}+e^{-i (\textbf{b}_1+\textbf{b}_2)\cdot\textbf{r}_0}\right)(\#\Gamma_{1}+\#\Gamma_{3})\nonumber\\
&&+i e^{-i (\textbf{b}_1+\textbf{b}_2)\cdot\textbf{r}_0} ([M_1]+[M_3])+i \left(e^{-i \textbf{b}_1\cdot\textbf{r}_0}+e^{-i \textbf{b}_2\cdot\textbf{r}_0}\right) [ X_1]\nonumber\\
&&-i \left(1+e^{-i \textbf{b}_1\cdot\textbf{r}_0}+e^{-i \textbf{b}_2\cdot\textbf{r}_0}+e^{-i (\textbf{b}_1+\textbf{b}_2)\cdot\textbf{r}_0}\right)(\#\Gamma_{2}+\#\Gamma_{4})\nonumber\\
&&-i e^{-i (\textbf{b}_1+\textbf{b}_2)\cdot\textbf{r}_0} ([M_2]+[M_4])-i\left(e^{-i \textbf{b}_1\cdot\textbf{r}_0}+e^{-i \textbf{b}_2\cdot\textbf{r}_0}\right) [ X_2]\nonumber\\
&=&2i\left\{ \left(\frac{1+e^{-i \textbf{b}_1\cdot\textbf{r}_0}+e^{-i \textbf{b}_2\cdot\textbf{r}_0}+e^{-i (\textbf{b}_1+\textbf{b}_2)\cdot\textbf{r}_0}}{2}\right)(\#\Gamma_{1}+\#\Gamma_{3}-\#\Gamma_{2}-\#\Gamma_{4})\right.\nonumber\\
&&\left.\hspace{4cm}+e^{-i (\textbf{b}_1+\textbf{b}_2)\cdot\textbf{r}_0} ([M_1]-[M_2])+\left(e^{-i \textbf{b}_1\cdot\textbf{r}_0}+e^{-i \textbf{b}_2\cdot\textbf{r}_0}\right) [ X]\right\},\nonumber
\end{eqnarray}
where we used $[X]\equiv[X_1]=-[X_2]$ due to particle-hole symmetry as well. By evaluating these expressions at the rotation centers $\textbf{r}_0=\textbf{a}_1,\textbf{a}_1+\textbf{a}_2$, we obtain
\begin{eqnarray}
T^{c(2)}_{\mathcal{S}}\vert_{\textbf{r}_0=\textbf{a}_1}&=&-2i\left([M_1]-[M_2]\right)\\
T^{c(2)}_{\mathcal{S}}\vert_{\textbf{r}_0=\textbf{a}_1+\textbf{a}_2}&=&2i\left\{  ([M_1]-[M_2])-2 [ X]\right\}\\
T^{c(4)}_{\mathcal{S}}\vert_{\textbf{r}_0=\textbf{a}_1+\textbf{a}_2}&=&-i\sqrt{2}\left([M_1]+[M_2]\right),
\end{eqnarray}
which finally implies, using Eq.\ref{Eq_rot}, that
\begin{eqnarray}
[X]&=&\frac{i}{4}\left(T^{c(2)}_{\mathcal{S}}\vert_{\textbf{r}_0=\textbf{a}_1}+T^{c(2)}_{\mathcal{S}}\vert_{\textbf{r}_0=\textbf{a}_1+\textbf{a}_2}\right)=\frac{i}{4}\text{Tr}\left[\left(\overline{c}^{(2)}_{\mathcal{S}}|_{\textbf{r}_0=\mathbf{a}_1}\right)+\left(\bar{c}^{(2)}_{\mathcal{S}}|_{\textbf{r}_0=\textbf{a}_1+\textbf{a}_2}\right)\right],\\
\left[M_{1,2}\right]&=&\frac{i}{4}\left(\sqrt{2}T^{c(4)}_{\mathcal{S}}\vert_{\textbf{r}_0=\textbf{a}_1+\textbf{a}_2}\pm T^{c(2)}_{\mathcal{S}}\vert_{\textbf{r}_0=\textbf{a}_1}\right)=\frac{i}{4}\text{Tr}\left[\sqrt{2}\left(\bar{c}^{(4)}_{\mathcal{S}}|_{\textbf{r}_0=\mathbf{a}_1+\textbf{a}_2}\right)\pm \left(\bar{c}^{(2)}_{\mathcal{S}}|_{\textbf{r}_0=\mathbf{a}_1}\right)\right],
\end{eqnarray}
which are the expressions discussed in the main text. The topological marker for this invariant satisfies
\begin{eqnarray}
    \vert\mathcal{T}^{c(n)}_{\mathcal{S}}(\textbf{r})\vert &<&\mathcal{O}\left(e^{-2\sin\left[\frac{\pi}{n}\right]\vert \textbf{r}-\mathcal{S}\vert/\zeta}\right) \quad \text{when} \quad \zeta \ll \vert x-\mathcal{S}\vert,
\end{eqnarray}
and an impurity placed at $\textbf{r}=\textbf{R}_0$ will change an invariant centered at $\mathcal{S}$ by an exponentially small amount
\begin{equation}
   \vert T^{c(n)'}_{\mathcal{S}}(\textbf{R}_0)-T^{c(n)}_{\mathcal{S}}\vert<\mathcal{O}\left( e^{-2\sin\left[\frac{\pi}{n}\right]\vert \textbf{R}_0- \mathcal{S}\vert/\zeta}\right) \quad \text{when} \quad \zeta \ll \vert \textbf{R}_0-\mathcal{S}\vert,
\end{equation}
where the notation $\vert \mathbf{R}_0 -\mathcal{S} \vert$ means the shortest distance to any of the invariant points in the set $\mathcal{S}.$

\subsection{III.3 Model example}
To exemplify the topological markers associated with these invariants, we used the model
\begin{eqnarray}
h(\mathbf{k})=\sin(\mathbf{k}.\mathbf{a}_1)\tau_1+\sin(\mathbf{k}.\mathbf{a}_2)\tau_2+u_1\left[\cos(\mathbf{k}.\mathbf{a}_1)+\cos(\mathbf{k}.\mathbf{a}_2)\right]\tau_3 
+u_2\left[\cos(\mathbf{k}.\mathbf{a}'_1)+\cos(\mathbf{k}.\mathbf{a}'_2)\right]\tau_3,
\end{eqnarray}
where $\textbf{a}_1=a(1,0),$ $\textbf{a}_2=a(0,1),$ $\textbf{a}'_1=\textbf{a}_1+\textbf{a}_2,$ and $\textbf{a}'_1=-\textbf{a}_1+\textbf{a}_2.$ The Pauli matrices $\tau_a$ act on the Nambu degree of freedom. The four-fold rotation operator is given by $\widetilde{c}^{(4)}_{\mathcal{S}}(\textbf{k})\vert_{\textbf{r}_0=\textbf{0}}=\frac{1}{\sqrt{2}}\left(\tau_0+i\tau_3\right).$ The specific parameters chosen were $u_1=1.0$ and $u_2=0.5.$

%%%%%%%%%%%%%%%%%%%%%%%%%%%%%%%%%%%%%%%%%%%%%%%

\section{Section IV: Mirror-symmetric DIII topological superconductors in two dimensions}

\subsection{IV.1 Conventional classification}

In the main text, we discussed the basis-independent approach to the classification of mirror-symmetric DIII topological supercondctors in two dimensions. Here, we will make contact with the momentum-space classification in order to connect with Eq.\ref{Eq_gen}. 

As we mentioned in the main text, a 2D mirror-symmetric superconductor in class DIII is described by a Bogoliubov-de Gennes Hamiltonian $\hat{H}$ such that $[\hat{m}_{\mathcal{S}},\hat{H}]=0.$ The mirror-invariant lines are located at $\mathcal{S}=(Y_1+y_0,Y_2+y_0),$ with $y_0=0,1/2.$ We focus on the case when $\hat{m}^2_{\mathcal{S}}=-1.$ If we Fourier transform only the coordinate that changes under mirror symmetry, the mirror constraint on the Bloch Hamiltonian is
\begin{equation}
\widetilde{m}_{\mathcal{S}}(k_y) h(k_y) \widetilde{m}^{-1}_{\mathcal{S}}(k_y)=h(-k_y),   
\end{equation}
where $\widetilde{m}_{\mathcal{S}}(k_y)$  is the momentum representation of the mirror operator.  At the mirror invariant momentum lines $k^\text{inv}_{y}=0,\pi,$ the Bloch states can be labeled with mirror eigenvalues. By using the  chiral symmetry $\chi=TC$ that arises from having both time-reversal $T$ and particle-hole $C$ symmetries, one can define topologically invariant mirror-winding numbers given by \cite{MS2014}
\begin{equation}
    \nu_{\mathcal{S}}^{(\pm)}(k^\text{inv}_{y})= -\frac{1}{N_x}\text{Tr}\left[(S_-\mathcal{Q}_{\pm,k^{\text{inv}}_y}S_+) \left[\hat{X}, (S_+\mathcal{Q}_{\pm,k^{\text{inv}}_y}S_-)\right]\right],
\end{equation}
where $\mathcal{Q}_{\pm,k^{\text{inv}}_y}=2\mathcal{P}_{\pm,k^{\text{inv}}_y}-\mathbb{I},$ $S_{\pm}=(\mathbb{I}\pm \hat{\chi})/2$ projects into chiral subspaces with eigenvalue $\chi=\pm1,$ and $\mathcal{P}_{\pm,k^{\text{inv}}_y}$ is the projection operator into the occupied states with mirror eigenvalue $\pm i.$ Due to time-reversal symmetry, $\nu_{\mathcal{S}}^{(+)}(k^\text{inv}_{y})=-\nu_{\mathcal{S}}^{(-)}(k^\text{inv}_{y}).$ The integers at the two invariant momenta leads to the known $\mathbb{Z}\times \mathbb{Z}$ classification \cite{Zhang2013}. To express the winding in terms explicitly of projection operators, we write
\begin{eqnarray}
    \nu_{\mathcal{S}}^{(\pm)}(k^\text{inv}_{y})&=&-\frac{1}{N_x}\text{Tr}\left[(S_-\mathcal{Q}_{\pm,k^{\text{inv}}_y}S_+) \left[\hat{X}, (S_+\mathcal{Q}_{\pm,k^{\text{inv}}_y}S_-)\right]\right]\\
    &=&-\frac{1}{N_x}\text{Tr}\left[(2\mathcal{P}_{\pm,k^{\text{inv}}_y}-\mathbb{I}) S_+\left[\hat{X},(2\mathcal{P}_{\pm,k^{\text{inv}}_y}-\mathbb{I})\right]S_-\right]\\
    &=&-\frac{4}{N_x}\text{Tr}\left[\mathcal{P}_{\pm,k^{\text{inv}}_y} \left[\hat{X},S_+\mathcal{P}_{\pm,k^{\text{inv}}_y}S_-\right]\right]\\
    &=&-\frac{4}{N_x}\text{Tr}\left[\mathcal{P}_{\pm,k^{\text{inv}}_y} \left[\hat{X},\left(\frac{(\mathbb{I}+ \hat{\chi})}{2}\right)\mathcal{P}_{\pm,k^{\text{inv}}_y}\left(\frac{(\mathbb{I}- \hat{\chi})}{2}\right)\right]\right]\\
    &=&-\frac{1}{N_x}\text{Tr}\left[\mathcal{P}_{\pm,k^{\text{inv}}_y} \left[\hat{X},\mathcal{P}_{\pm,k^{\text{inv}}_y}+\hat{\chi}\mathcal{P}_{\pm,k^{\text{inv}}_y}-\mathcal{P}_{\pm,k^{\text{inv}}_y}\hat{\chi}-\hat{\chi}\mathcal{P}_{\pm,k^{\text{inv}}_y}\hat{\chi}\right]\right]\\
     &=&-\frac{1}{N_x}\text{Tr}\left[\mathcal{P}_{\pm,k^{\text{inv}}_y} \left[\hat{X},\mathcal{P}_{\pm,k^{\text{inv}}_y}+(\hat{\chi}-\mathcal{P}_{\pm,k^{\text{inv}}_y}\hat{\chi})-\mathcal{P}_{\pm,k^{\text{inv}}_y}\hat{\chi}-(\mathbb{I}-\mathcal{P}_{\pm,k^{\text{inv}}_y})\right]\right]\\
     &=&-\frac{2}{N_x}\text{Tr}\left[\mathcal{P}_{\pm,k^{\text{inv}}_y} \left[\hat{X},\mathcal{P}_{\pm,k^{\text{inv}}_y}\right]\right]+\frac{2}{N_x}\text{Tr}\left[\mathcal{P}_{\pm,k^{\text{inv}}_y} \left[\hat{X},\mathcal{P}_{\pm,k^{\text{inv}}_y}\hat{\chi}\right]\right]\\
    &=&\frac{2}{N_x}\text{Tr}\left[\mathcal{P}_{\pm,k^{\text{inv}}_y} \left[\hat{X}, \mathcal{P}_{\pm,k^{\text{inv}}_y}\right] \hat{\chi}\right], \label{Eq_nuk}
\end{eqnarray}
which is the expression we presented in the main text.

\subsection{IV.2 Basis-independent classification}
We now implement Eq.\ref{Eq_gen} for this TCP. As is clear from the previous section, the topological invariants of the invariant subspaces are
\begin{equation}
    \tau^{\pm}_{\mathcal{S},b}(k^{\text{inv}}_y)=\nu^{\pm}_{\mathcal{S}}(k^{\text{inv}}_y),   
\end{equation}
where we added the index $b$ to be reminded that it refers to the bulk invariant, as opposed to the edge invariant that we introduce in the next subsection. Furthermore, we have that
\begin{equation}
    \mathcal{F}(\mathcal{P})=2N^{-1}_x\mathcal{P}\left[\hat{X},\mathcal{P}\right]\chi.
\end{equation} By applying Eq.\ref{Eq_gen}, we then have the topological invariants
\begin{eqnarray}
T^{m}_{\mathcal{S}}=\sum_{k^{\text{inv}}_y}\left[(+i)\nu^{+}_{\mathcal{S}}(k^{\text{inv}}_y)+(-i)\nu^{-}_{\mathcal{S}}(k^{\text{inv}}_y)\right]= \text{Tr}\left[\overline{m}_{\mathcal{S}}\mathcal{F}(\mathcal{P}_G)\right].
\end{eqnarray}
The relation with the invariant we defined in the main text is then
\begin{equation}
    \mathcal{V}^m_{\mathcal{S}}=-\frac{i}{2} T^{m}_{\mathcal{S}} \label{Eq_locnu}
\end{equation}
The topological marker for this invariant satisfies
\begin{eqnarray}
    \vert\mathcal{T}^{m}_{\mathcal{S}}(y)\vert &<&\mathcal{O}\left(e^{-2\vert y-\mathcal{S}\vert/\zeta}\right) \quad \text{when} \quad \zeta \ll \vert y-\mathcal{S}\vert,
\end{eqnarray}
and an impurity placed at $y=Y_0$ will change an invariant centered at $\mathcal{S}$ by an exponentially small amount
\begin{equation}
   \vert T^{m'}_{\mathcal{S}}(Y_0)-T^{m}_{\mathcal{S}}\vert<\mathcal{O}\left( e^{-2\vert Y_0- \mathcal{S}\vert/\zeta}\right) \quad \text{when} \quad \zeta \ll \vert Y_0-\mathcal{S}\vert.
\end{equation}

\subsection{IV.3  Integer topological invariants for the edge}

In the main text, we used the bulk-boundary relation $\Delta^{m}_{\mathcal{S}}=-\frac{i}{2} \text{Tr}\left[\overline{m}^{edge}_\mathcal{S}\chi\right]=\mathcal{V}^{m}_{\mathcal{S}},$ which we will now prove in this section. To do this, we begin by noting that we can write $\Delta^{m}_{\mathcal{S}}$ in the suggestive form
\begin{eqnarray}
\Delta^{m}_{\mathcal{S}}&=&-\frac{i}{2} \text{Tr}\left[\overline{m}_\mathcal{S}\mathcal{F}\left(\mathcal{P}_{edge}\right)\right],
\end{eqnarray}
which has the form of the invariants in Eq.\ref{Eq_gen}, in this case with the function $\mathcal{F}\left(\mathcal{P}\right)=\mathcal{P}\chi.$ Thus, in momentum space we must obtain that $\Delta^{m}_{\mathcal{S}}$ receives contribution exclusively from the invariant momenta in the form
\begin{eqnarray}
\Delta^{m}_{\mathcal{S}}&=&\left(-\frac{i}{2}\right)\sum_{k^{\text{inv}}_y}\left[(+i)\tau^{edge}_{+}(k^{\text{inv}}_y)+(-i)\tau^{edge}_{-}(k^{\text{inv}}_y)\right],\label{Eq_DeltaTau}
\end{eqnarray}
where we defined new \textit{edge} topological invariants
\begin{equation}
    \tau^{edge}_{m}(k^{\text{inv}}_y)\equiv\text{Tr}\left[\mathcal{P}_{edge}(k^{\text{inv}}_y)\mathcal{F}\left(\mathcal{P}_{edge}(k^{\text{inv}}_y)\right)\right]=\text{Tr}\left[\mathcal{P}_{edge}(k^{\text{inv}}_y)\chi\right],
\end{equation} 
with $\mathcal{P}_{edge}(k^{\text{inv}}_y)$ the projector onto the boundary states with momentum $k^{\text{inv}}_y$ and mirror eigenvalue $m \cdot i.$ Because of the form of the function $\mathcal{F}$ in this case, the quantities $\tau^{edge}_{m}(k^{\text{inv}}_y)$ count the difference
\begin{equation}
    \text{Tr}\left[\mathcal{P}_{edge}(k^{\text{inv}}_y)\chi\right]=\sum_{n \in \text{edge}}\bra{u^{n}_{k^{\text{inv}}_y}}\chi\ket{u^n_{k^{\text{inv}}_y}}=\mathcal{N}^{m}_{+}(k^{\text{inv}}_y)-\mathcal{N}^{m}_{-}(k^{\text{inv}}_y)
\end{equation}
at the invariant momentum $k^{\text{inv}}_y$ in a given mirror subspace $m,$ with $\mathcal{N}^{m}_{\chi}(k^{\text{inv}}_y)$ the number of modes with chiral eigenvalue $\chi=\pm 1.$ Due to the bulk-boundary correspondence of AIII systems, this difference must be equal to the bulk topological winding Eq.\ref{Eq_nuk}, and thus
\begin{equation}
    \tau^{edge}_{m}(k^{\text{inv}}_y)=\nu^{(m)}_{\mathcal{S}}(k^{\text{inv}}_y).\label{Eq_bulkedge}
\end{equation}
Finally, putting together Eq.\ref{Eq_locnu}, Eq.\ref{Eq_DeltaTau} and Eq.\ref{Eq_bulkedge}, we obtain the relation
\begin{eqnarray}
\Delta^{m}_{\mathcal{S}}=\frac{1}{2}\sum_{k^{\text{inv}}_y}\left[\tau^{edge}_{+}(k^{\text{inv}}_y)-\tau^{edge}_{-}(k^{\text{inv}}_y)\right]=\frac{1}{2}\sum_{k^{\text{inv}}_y}\left[\nu^{(+)}_{\mathcal{S}}(k^{\text{inv}}_y)-\nu^{(-)}_{\mathcal{S}}(k^{\text{inv}}_y)\right]=\mathcal{V}^{m}_{\mathcal{S}}.
\end{eqnarray}
as we had set out to prove.

\subsection{IV.4 Quantized edge shift of gapless Majorana boundary modes}

\noindent In this section, we derive Eq.7 from the main text, which relates the local edge shift $\delta\mathcal{Y}^{\chi}_{edge}(R_1)$ with the bulk invariant $\mathcal{V}^{m}_{\mathcal{S}}.$  Consider the projected position operator $\mathcal{W}^{\chi}_{edge}=\mathcal{P}^{\chi}_{edge} e^{-2\pi i \hat{Y}/L_y} \mathcal{P}^{\chi}_{edge}.$ We denote its eigenvalues and eigenstates by $\{e^{-2\pi i \lambda_{R_yj}/L_y}\}$ and $\{\ket{\Psi^{\chi}_{R_yj}}\},$ respectively. Here, $R_y$ labels unit cells along the edge, and $j=1,\ldots, N_b,$ where $N_b$ is the number of subbands included in the detached Majorana band (DMB). Thus, the number of states in the edge projector $\mathcal{P}_{edge}$ is $N_b L_y.$ Let us write $\lambda_{R_yj}=R_y+\nu_{R_yj}.$ The edge shift is then defined by
\begin{equation}
    \delta\mathcal{Y}^{\chi}_{edge}(R_1)=\left(\sum^{N_b}_{j=1}\nu_{R_1j} \right)\text{mod}1.
\end{equation}
Now, mirror symmetry imposes the constraint
\begin{equation}
    \hat{m}_{\mathcal{S}}\mathcal{W}^{\chi}_{edge}\hat{m}^{-1}_{\mathcal{S}}=\mathcal{P}^{\chi}_{edge} \hat{m}_{\mathcal{S}} e^{-2\pi i \hat{Y}/L_y}\hat{m}^{-1}_{\mathcal{S}} \mathcal{P}^{\chi}_{edge}=e^{-4\pi iS/L_y}\left(\mathcal{W}^{\chi}_{edge}\right)^{\dagger}.
\end{equation}
Thus, we have that
\begin{equation}
    \mathcal{W}^{\chi}_{edge}\ket{\Psi^{\chi}_{R_yj}}=e^{-2\pi i\lambda_{R_yj}/L_y}\ket{\Psi^{\chi}_{R_yj}} \quad \longrightarrow \quad \mathcal{W}^{\chi}_{edge}\hat{m}_{\mathcal{S}}\ket{\Psi^{\chi}_{R_yj}}=e^{-2\pi i (2S-\lambda_{R_yj})/L_y}\hat{m}_{\mathcal{S}}\ket{\Psi^{\chi}_{R_yj}}.
\end{equation}
Because of this constraint, for a given fixed unit cell $R_y,$ there are three types of Wannier centers $\nu_{R_yj}:$
\begin{enumerate}
    \item Those that come in symmetry-related pairs $\{\nu_{R_yj} \text{mod}1,-\nu_{R_yj}\text{mod}1\}.$
    \item Those that coincide with an integer mirror-invariant point $\nu_{R_yj}\text{mod}1=0.$
    \item Those that coincide with a half-integer mirror-invariant point $\nu_{R_yj}\text{mod}1=1/2.$
\end{enumerate}
In view of these possibilities, if we denote by $N_{\chi,1/2}(R_1)$ the number of Wannier states centered at $R_1+1/2,$ then
\begin{eqnarray}
    \delta\mathcal{Y}^{\chi}_{edge}(R_1)=\left(\sum^{N_b}_{j=1}\lambda_{R_1 j} \right)\text{mod}1=\left(\sum^{N_{\chi,1/2}}_{i=1}\frac{1}{2}\right)\text{mod}1=\left[\frac{1}{2}N_{\chi,1/2}(R_1)\right]\text{mod}1.
\end{eqnarray}
Thus, the only way to obtain a nonzero shift $\delta\mathcal{Y}^{\chi}_{edge}(R_1)=1/2$ is for there to be an odd number of Wannier centers at the mid-point of the unit cell $R_1.$ Next, let us write $N_{\chi,1/2}(R_1)=N^{(+)}_{\chi,1/2}(R_1)+N^{(-)}_{\chi,1/2}(R_1),$ where $N^{(\pm)}_{\chi,1/2}(R_1)$ is the number of Wannier states centered at $R_1+1/2$ with $\pm i$ mirror eigenvalues. Then
\begin{eqnarray}
    \delta\mathcal{Y}^{\chi}_{edge}(R_1)=\left[\frac{1}{2}\left(N^{(+)}_{\chi,1/2}(R_1)+N^{(-)}_{\chi,1/2}(R_1)\right)\right]\text{mod}1=\left[\frac{1}{2}\left(N^{(+)}_{\chi,1/2}(R_1)-N^{(-)}_{\chi,1/2}(R_1)\right)\right]\text{mod}1. \label{Eq_dYndiff}
\end{eqnarray}
Consider the set of mirror-invariant points $S\vert_{y_0=1/2}=\{R_1+1/2,R_2+1/2\}.$ Due to the constraint of mirror symmetry, for $\lambda_{R_yj} \neq R_1+1/2,$ the state $\ket{\Psi^{\chi}_{R_yj}}$ is orthogonal to the state $\hat{m}_{\mathcal{S}}\ket{\Psi^{\chi}_{R_yj}}.$  By contrast, if $\lambda_{R_yj}=R_1+1/2,$ then $\ket{\Psi^{\chi}_{R_yj}}$ is an eigenstate of $\hat{m}_{\mathcal{S}}\vert_{y_0=1/2}.$ It follows that
\begin{equation}
    N^{(+)}_{\chi,1/2}(R_1)-N^{(-)}_{\chi,1/2}(R_1)=-\frac{i}{2}\sum_{R_yj}\bra{\Psi^{\chi}_{R_yj}}(\hat{m}_{\mathcal{S}}\vert_{y_0=1/2})\ket{\Psi^{\chi}_{R_yj}}.\label{Eq_Ndiff}
\end{equation}
Note that the dependence of the right-hand side on $R_1$ comes from the index $S\vert_{y_0=1/2}$ of the mirror operator that is chosen. The factor of $1/2$ is due to the fact that the sum counts the \textit{two} points in $S\vert_{y_0=1/2},$ and we defined $N_{\chi,1/2}$ with respect to only \textit{one} of them. Using Eq.\ref{Eq_Ndiff} in Eq.\ref{Eq_dYndiff}, we then obain
\begin{eqnarray}
    \delta\mathcal{Y}^{\chi}_{edge}(R_1)&=&\left[-\frac{i}{4}\sum_{R_yj}\bra{\Psi^{\chi}_{R_yj}}(\hat{m}_{\mathcal{S}}\vert_{y_0=1/2})\ket{\Psi^{\chi}_{R_yj}}\right] \text{mod}\, 1.
\end{eqnarray}
Now, note that since $\hat{\chi}T\ket{\Psi^{+}_{R_yj}}=-\chi T\ket{\Psi^{+}_{R_yj}}$ then $T\ket{\Psi^{+}_{R_yj}}\propto\ket{\Psi^{-}_{R_yj}}.$ Thus if $\hat{m}_{\mathcal{S}}\vert_{y_0=1/2}\ket{\Psi^{+}_{R_yj}}=e^{i\phi}\ket{\Psi^{+}_{R_yj}},$ then by acting with $T$ on both sides we obtain $\hat{m}_{\mathcal{S}}\vert_{y_0=1/2}\ket{\Psi^{-}_{R_yj}}=e^{-i\phi}\ket{\Psi^{-}_{R_yj}}$ i.e. the two states $\ket{\Psi^{+}_{R_yj}}$ and $\ket{\Psi^{-}_{R_yj}}$ have opposite mirror eigenvalues. Because of this, we can write
\begin{equation}
    \bra{\Psi^{\chi}_{R_yj}}(\hat{m}_{\mathcal{S}}\vert_{y_0=1/2})\ket{\Psi^{\chi}_{R_yj}}=\frac{1}{2}\left(\bra{\Psi^{+}_{R_yj}}(\hat{m}_{\mathcal{S}}\vert_{y_0=1/2})\ket{\Psi^{+}_{R_yj}}-\bra{\Psi^{-}_{R_yj}}(\hat{m}_{\mathcal{S}}\vert_{y_0=1/2})\ket{\Psi^{-}_{R_yj}}\right).
\end{equation}
We then obtain
\begin{eqnarray}
    \delta\mathcal{Y}^{\chi}_{edge}(R_1)&=&\left[-\frac{i}{8}\sum_{R_yj}\left\{\bra{\Psi^{+}_{R_yj}}(\hat{m}_{\mathcal{S}}\vert_{y_0=1/2})\ket{\Psi^{+}_{R_yj}}-\bra{\Psi^{-}_{R_yj}}(\hat{m}_{\mathcal{S}}\vert_{y_0=1/2})\ket{\Psi^{-}_{R_yj}}\right\}\right] \text{mod}\, 1,
\end{eqnarray}
Finally, we replace the sum over Wannier functions by a trace over the full Hilbert space of the system by introducing the projection operator $\mathcal{P}_{edge}:$
\begin{eqnarray}
    \delta\mathcal{Y}^{\chi}_{edge}(R_1)&=&\left[-\frac{i}{8}\sum_{R_yj\chi}\bra{\Psi^{\chi}_{R_yj}}(\hat{m}_{\mathcal{S}}\vert_{y_0=1/2})\hat{\chi}\ket{\Psi^{\chi}_{R_yj}}\right] \text{mod}\, 1,\\
    &=&\left[-\frac{i}{8}\text{Tr}\left[\mathcal{P}_{edge}(\hat{m}_{\mathcal{S}}\vert_{y_0=1/2})\hat{\chi}\right]\right] \text{mod}\, 1,\\
    &=&\left[\frac{1}{4}\Delta^m_{\mathcal{S}}\vert_{y_0=1/2}\right] \text{mod}\, 1,\\
    &=&\left[\frac{1}{4}\mathcal{V}^m_{\mathcal{S}}\vert_{y_0=1/2}\right] \text{mod}\, 1,
\end{eqnarray}
which is the expression we discussed in the main text.

\subsection{IV.5 Model example}
We here provide the details of the particular model we used to illustrate the bulk and edge topology of the two DIII-trivial topological phases that were discussed in the main text. The model we used was designed to expose clearly the features that arise in these phases, particularly at the edge. The two models we discussed had the topological invariants
\begin{eqnarray}
    \text{Weak TCP}:&& \mathcal{V}^{m}_{\mathcal{S}}\vert_{y_0=0}=2, \quad \mathcal{V}^{m}_{\mathcal{S}}\vert_{y_0=1/2}=0 \label{Eq_wtop}\\ \text{Mirror-topological TCP}:&& \mathcal{V}^{m}_{\mathcal{S}}\vert_{y_0=0}=0, \quad \mathcal{V}^{m}_{\mathcal{S}}\vert_{y_0=1/2}=2. \label{Eq_mtop}
\end{eqnarray}

This is achieved by the Hamiltonian that is effectively two copies of the model from \cite{Zhang2009}:
\begin{equation}
h({\bf{k}})=\sin k_x \Gamma_1 + t_y\sin k_y \left(\Gamma_2+m_3 \Lambda_3\right)+
\left(\mu+\cos k_x + t_y \cos k_y\right) \Gamma_3+ \sum^2_{i=1} m_i \Lambda_i,
\end{equation}
where
\begin{equation}
    \Gamma_1=f_3 \tau_3 \sigma_3 , \hspace{0.5cm} \Gamma_2=f_0 \tau_3 \sigma_1 ,  \hspace{0.5cm} \Gamma_3= f_0  \tau_1 \sigma_0 , \hspace{0.5cm} \Lambda_1=f_3 \tau_1 \sigma_0  , \hspace{0.5cm}\Lambda_2=f_2 \tau_1 \sigma_3  \hspace{0.5cm} \Lambda_3=f_1\tau_3\sigma_2 .
\end{equation}
The Pauli operators $\tau_a,$ $\sigma_a$ act on Nambu and spin degrees of freedom, and $f_a$ acts on the two copies. The time-reversal, charge conjugation, chiral, and mirror operators are $T=\sigma_2 K,$ $C=i \tau_2\sigma_2 K,$ $S=\tau_2,$ and $\widetilde{m}_{\mathcal{S}}(k_y)=ie^{2ik_y y_0}\sigma_3$ respectively. The parameters we used to achieve the two phases are:
\begin{center}
\begin{tabular}{ |c|c|c|c|c|c| } 
 \hline
  & $t_y$ & $\mu$ & $m_1$ & $m_2$ & $m_3$\\ 
 \hline
 Weak TCP & $0.25$ & $-1.0$ & $1.0$ & $0.0$ & $0.0$\\ 
 \hline
 Mirror-topological TCP & $1.0$ & $0.0$ & $-1.0$ & $0.5$ & $0.5$ \\
 \hline
\end{tabular}
\end{center}
To check that the invariants yield the correct values, we can simply evaluate the Bloch Hamiltonian at the invariant momenta $k^{\text{inv}}_y.$ When $m_2=0,$ the Bloch Hamiltonian is block-diagonal in the eigenstates of $\widetilde{m}_{\mathcal{S}}(k_y).$ By denoting $h^{(\pm)}(k_x,k^{\text{inv}}_y)$ the diagonal blocks, then the positive mirror block is 
\begin{eqnarray}
h^{(+)}(k_x,k^{\text{inv}}_y)&=& \sin k_x f_3 \tau_3+
\left(\mu+\cos k_x+ t_y e^{i k^{\text{inv}}_y}\right) f_0\tau_1+ m_1 f_3 \tau_1,\\
&=&\left( \begin{array}{cc}
 \sin k_x\tau_3+\left(M_{+}(k^{\text{inv}}_y)+\cos k_x\right) \tau_1 & 0 \\
0 &  -\sin k_x\tau_3+\left(M_{-}(k^{\text{inv}}_y)+\cos k_x\right) \tau_1 
\end{array} \right)
\end{eqnarray}
where $M_{\pm}(k^{\text{inv}}_y)=\mu+ t_y e^{i k^{\text{inv}}_y}\pm m_1.$ Each block of $h^{(+)}(k_x,k^{\text{inv}}_y)$ thus realizes a minimal two-band model of AIII topological wires. It is then straightforward to check that the momentum-space invariants for the weak TCP are $\nu^{+}_{\mathcal{S}}(0)=\nu^{+}_{\mathcal{S}}(\pi)=1,$ which leads to Eq.\ref{Eq_wtop}. Similarly, one obtains for the mirror-topological TCP $\nu^{+}_{\mathcal{S}}(0)=-\nu^{+}_{\mathcal{S}}(\pi)=1,$ which leads to Eq.\ref{Eq_mtop}. The added terms proportional to $m_{2,3}$ do not close the bulk gap and are chosen so as to detach the boundary modes from the bulk bands.

Finally, in order to numerically obtain the edge projector $\mathcal{P}_{edge}$ of the boundary bands of \textit{only one} of the edges, it is useful to gap out the boundary modes on the opposite edge to avoid any hybridization between both edge modes. To achieve this we added the following terms to the boundaries of the weak TCP
\begin{eqnarray}
V_{1}&=&(-1.0)\left(f_1\tau_3\sigma_0\right),\\
V_{L_x}&=&(0.5)\left(f_1\tau_3\sigma_1-\cos k_y f_0\tau_2\sigma_0\right),
\end{eqnarray}
and for the mirror-topological TCP we added
\begin{eqnarray}
V_{1}&=&(0.1)\left(f_1\tau_3\sigma_0\right),\\
V_{L_x}&=&(-0.8)\left(f_0\tau_3\sigma_1\right),
\end{eqnarray}
where $\bra{x,y}V_1\ket{x,y}=\delta_{x,1}$ and $\bra{x,y}V_{L_x}\ket{x,y}=\delta_{x,L_x}.$ This leaves gapless the boundary modes at the $x=1$ edge. These are the boundary modes for which the topological markers are shown in Fig.3 of the main text for both phases.

\end{document}